\documentclass[traditabstract]{aa} 
\usepackage{natbib}
\usepackage{graphicx}
\usepackage{txfonts}
\usepackage{multirow}
\usepackage{wasysym}
\usepackage{stmaryrd}
\usepackage{pifont}
\usepackage{gplPts}
\usepackage{mynewfonts}
\begin{document}
   \title{Deep U-B-V imaging of the Lockman Hole with the LBT\thanks{
Based on data acquired using the Large Binocular Telescope (LBT). The
LBT is an international collaboration among institutions in the United States,
Italy, and Germany. LBT Corporation partners are the University of Arizona
on behalf of the Arizona university system; Istituto Nazionale di Astrofisica,
Italy; LBT Beteiligungsgesellschaft, Germany, representing the Max-Planck
Society, the Astrophysical Institute Potsdam, and Heidelberg University; Ohio
State University, and the Research Corporation, on behalf of the University
of Notre Dame, the University of Minnesota, and the University of Virginia
}}

   \subtitle{Observations and number counts}

   \author{E. Rovilos\inst{1}\fnmsep\thanks{send off-print requests to erovilos@mpe.mpg.de}
           \and
           V. Burwitz\inst{1}
           \and
           G. Szokoly\inst{1,2}
           \and
           G. Hasinger\inst{3}
           \and
           E. Egami\inst{4}
           \and
           N. Bouch\'{e}\inst{1}
           \and
           S. Berta\inst{1}
           \and
           M. Salvato\inst{3,5}
           \and
           D. Lutz\inst{1}
           \and
           R. Genzel\inst{1}
          }
   \institute{Max Planck Institut f\"{u}r extraterrestrische Physik,
              Giessenbachstra\ss e, 85748 Garching, Germany\\
              \and
              Institute of Physics, E\"{o}tv\"{o}s University, P\'{a}zm\'{a}ny P. s. 1/A,
              1117 Budapest, Hungary\\
              \and
              Max Planck Institut f\"{u}r Plasmaphysik, Boltzmannstra\ss e 2, 85748, Garching, Germany\\
              \and
              Steward Observatory, University of Arizona, 933 North Cherry Avenue, Tucson,
              AZ 85721, USA\\
              \and
              California Institute of Technology, MC 205-24, 1200 East California Boulevard, Pasadena, CA 91125, USA\\
             }

   \date{Received ...; accepted ...}

 \abstract{
  We used the large binocular camera (LBC) mounted on the large binocular
  telescope (LBT) to observe the Lockman Hole in the U, B, and V bands. Our
  observations cover an area of 925\,arcmin$^2$. We reached depths of 26.7,
  26.3, and 26.3\,mag(AB) in the three bands, respectively, in terms of 50\%
  source detection efficiency, making this survey the deepest U-band survey
  and one of the deepest B and V band surveys with respect to its covered area.
  We extracted a large number of sources ($\sim89000$), detected
  in all three bands
  and examined their surface density, comparing it with models of galaxy
  evolution. We find good agreement with previous claims of a steep faint-end
  slope of the luminosity functions, caused by late-type and irregular
  galaxies at $z>1.5$. A population of dwarf star-forming galaxies at
  $1.5<z<2.5$ is needed to explain the U-band number counts. We also find
  evidence of strong supernova feedback at high redshift.
  This survey is complementary to the r, i, and z Lockman Hole survey conducted
  with the Subaru telescope and provides the essential wavelength coverage
  to derive photometric redshifts and select different types of sources from
  the Lockman Hole for further study.
  }{}{}{}{}

   \keywords{Surveys -- Galaxies: photometry}

   \maketitle
%

\section{Introduction}

The formation and evolution of cosmic structures, such as galaxies, clusters,
and the large-scale structure, are some of the most important issues in modern
astrophysics. According to hierarchical models, initial fluctuations of the
dark matter mass density develop to form galaxies, clusters, and
the cosmic web. Such processes leave their footprints in different regimes
of the electromagnetic spectrum, and assembling statistically significant
samples of extragalactic objects at different wavelengths can give valuable
information on the various processes involved in the evolution of the
universe.

A very valuable tool for constructing such samples is deep ``blind'' surveys,
where a region in the sky with no bright sources is observed with a long
integration time. Optical surveys are very important in this context, as
they are able to provide the densest fields in terms of detected sources
and serve as ``anchor points'' for the multi-wavelength coverage. After
a multi-wavelength coverage has been achieved, one could apply photometric
redshift techniques \citep*[e.g.][]{Bolzonella2000,Benitez2000,Ilbert2009}
to examine the luminosities of the various sources or select source samples
for spectroscopy.

Notable results have been reported in various fields of extragalactic
astrophysics using blind deep surveys. Combining imaging and
spectroscopic surveys at different regimes of the spectrum, different groups
have been able to derive the star formation \citep[e.g.][]{Hopkins2004} and
accretion histories \citep[e.g.][]{Ueda2003} of the universe and examine their
co-evolution \citep*{Vollmer2008,Somerville2008}.
From optical imaging and photometry
alone, one can use the information in the number count of the detected sources
to test the geometry and evolutionary models of the universe. For example,
Eucledian geometry would result in a constant slope of 0.6 in the galaxy
number counts with respect to their magnitudes, but this has been ruled out
from early results in this direction \citep*[e.g.][]{Gardner1993}. Measuring
the number counts in different wavebands, it is evident that simple geometric
models invoking a ``deceleration parameter'' (q) could not give good fits and
some kind of evolution has to be taken into account \citep{Metcalfe1991}.
This effect is more severe in blue colours in the form of excess counts at
fainter magnitudes and it is widely known as the ``faint blue galaxy problem''.
With high resolution observations using the HST, \citet{Driver1995}
demonstrate that the sources responsible for the faint counts have late-type and
irregular morphologies; adding a population of $z\simeq2$ dwarf star-forming
galaxies \citep{Metcalfe1995} gives a reasonable fit to the blue number counts
data. These galaxies contribute to the star formation at redshifts $z\gtrsim1$
and are merged or simply have evolved to non activity locally.

Support for this scenario comes from the study of the (blue) luminosity
functions of different kinds of objects at different redshifts.
\citet{Ilbert2005} find that bluer luminosity functions show evidence of
more rapid evolution with redshift than redder ones, and later spectral types
and bluer colours seem to play a more important role in it
\citep{Zucca2006,Willmer2006}. However, small evolution of the disc
population to $z\simeq1$ is observed by \citet{Ilbert2006}, but the strong
evolution of bulge-dominated systems could be attributed to a dwarf galaxy
population \citep[see also][]{Im2001}. The study luminosity function is
limited to relatively bright objects as it is based on redshifts. A number
count distribution can probe fainter objects and give an approximation on
the faint-end slope \citep{Barro2009} of the LF and help distinguish between
different results (see comparisons in \citealt{Ilbert2005} and
\citealt{Zucca2006}). In this paper we present deep U-B-V band observations
of the Lockman Hole with the corresponding number counts to 27.5\,mag(AB).

\section{The Lockman Hole multi-wavelength survey}

The Lockman Hole is a region with minimal galactic absorption
\citep*[$N_{\rm HI}=4.5\times10^{19}\,{\rm cm}^{-2}$,][]{Lockman1986}
and the absolute minimum of infrared cirrus emission in the sky.
Its position in the northern sky ($\alpha=10^{\rm h}52^{\rm m}43^{\rm s}$,
$\delta=57^{\circ}28\arcmin48\arcsec$) makes it an ideal location for deep
surveys. Indeed it has a large multi-wavelength coverage spanning from
X-rays to meter-wavelength radio. In X-rays it has been observed with the
ROSAT satellite \citep{Hasinger1998} and more recently with XMM
\citep{Hasinger2001,Brunner2008}, reaching a depth of
$1.9\times10^{-16}\,{\rm erg}\,{\rm cm}^{-2}\,{\rm s}^{-1}$ in the 0.5-2.0\,keV
band. In the
ultra-violet it has been observed by GALEX \citep{Martin2005} as one of
its deep fields, with the data being publically available. In the near
infrared (J and K bands) it is a part of the UKIDSS ultra deep survey
\citep{Lawrence2007} reaching K$\sim$23(AB). In infrared wavelengths it was
observed by ISO using both ISOPHOT and ISOCAM
\citep{Kawara2004,Fadda2004,Rodighiero2004} and more recently there have been
observations with Spitzer-IRAC \citep{Huang2004} and
Spitzer MIPS \citep{Egami2008}. The Lockman Hole is also part of
the SWIRE survey \citep{Lonsdale2003}, observed with both IRAC and MIPS and
covering a much wider (but shallower) area.
There have been a number of millimeter - sub-mm observations of the Lockman
Hole, namely with the JCMT-SCUBA \citep{Coppin2006}, JCMT-AzTEC
\citep{Scott2006}, IRAM-MAMBO \citep{Greve2004}, and CSO-Bolocam
\citep{Laurent2005}. In the radio regime, the Lockman Hole has been observed
with the VLA, both in 5 and in 1.4\,GHz \citep{Ciliegi2003,Ivison2002,Biggs2006}
and with MERLIN in 1.4\,GHz \citep{Biggs2008}. Finally, in meter-wavelengths
it was targeted by the GMRT \citep{Garn2008}.

In this work we present the results of an imaging campaign of the Lockman Hole
in the optical. We have used the LBT to obtain deep U, B, and V images.
The ``red'' part of the optical imaging campaign has been conducted with
the Subaru telescope (in the r, i, and z bands) and will be presented by
Szokoly et al. (in preparation).


\section{Observations}

The observations were made with the Large Binocular Camera
\citep[LBC,][]{Giallongo2008} of the Large Binocular Telescope (LBT) on Mount
Graham, Arizona.
The LBT has two 8.4\,m mirrors on a common mount and both of them are
equipped with a prime focus camera. Both LBCs contain four CCD chips with
2048$\times$4608\,pixels each. Three chips are aligned parallel to each other
while the fourth is tilted by 90 degrees and located above them.
This provides a 23$\times$23\,arcmin field of view with a sampling of
0.23\,arcsec/pixel.
The gaps between the CCDs are 945\,nm wide, which corresponds to 18\,arcsec,
thus a 5-point circular dither pattern with a diameter of 30\,arcsec was
chosen to provide good coverage over the whole area.

\begin{figure}
  \resizebox{\hsize}{!}{\includegraphics{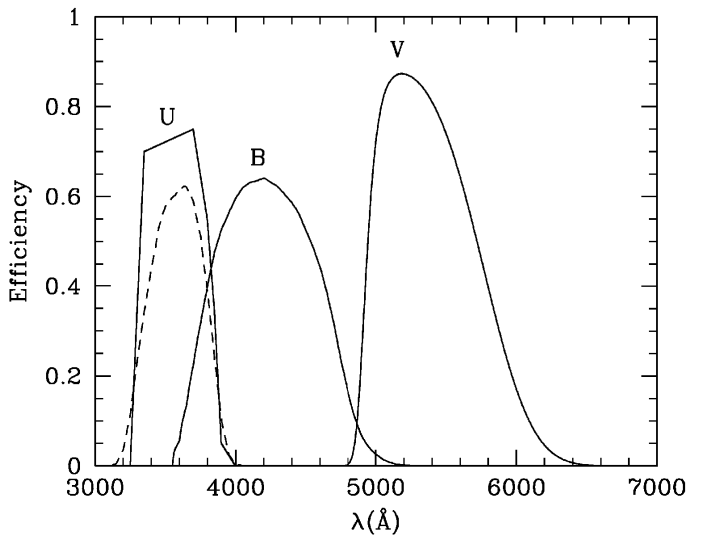}}
  \caption{Transmission curves of the filters used (U-spec - B-Bessel -
           V-Bessel)
           are shown by continuous lines. The dotted
           line represents the U-Bessel filter available on the LBC. We have
           chosen the U-spec filter for the U-band on grounds of its better
           efficiency and more uniform spectral coverage. These curves
           represent the filters' responses without accounting for the
           detectors' responses or the atmosphere. The detectors' responses are
           slightly different for the two arms of the telescope which might have
           an effect in the V-band.}
  \label{filters}
\end{figure}

Both cameras have an 8 position filter wheel each, and together a total of 13
filters are available,
covering a range from the ultraviolet to the near-infrared. For the U-band
Lockman Hole imaging we used the special LBT U-band filter (see Fig.
\ref{filters}) which has a more uniform coverage and better efficiency than
the standard U-Bessel. For the other images we used the
standard B-Bessel and V-Bessel filters.

The Lockman Hole was observed in March, April and May 2007 during
science demonstration time (SDT; PI: E. Egami), when only the camera on the
``blue'' channel of the telescope was available, and in 2008 and 2009 during
LBTB (German institutes') time (PI: G. Hasinger) with both cameras
available in ``binocular'' mode.
We have chosen 2 different pointings as centres of the image, corresponding
to the VLA ($\alpha=10^{\rm h}52^{\rm m}08.8^{\rm s}$,
$\delta=57^{\circ}21\arcmin34\arcsec$, \citealt{Ivison2002}) and the XMM
($\alpha=10^{\rm h}52^{\rm m}43^{\rm s}$, $\delta=57^{\circ}28\arcmin48\arcsec$,
\citealt{Hasinger2001})
pointings. These are separated by 8.6\,arcmin, so there is a large area of
overlap, where our images have
the highest sensitivity. During the science demonstration time the observing
time was split in half between XMM and VLA exposures and during the LBTB
time we concentrated on the XMM area.

The total time spent on the Lockman Hole was 36.8 hours which are distributed
among the various filters using both channels (when available) as described
in Tab. \ref{observations}. The exposure time for each observation was 360
seconds initially, but it was reduced to 180 seconds for later observing
runs (May 2008 onwards), after discovering a large
number of saturated sources and limited source tracking efficiency of the
telescope for long exposures. The effective exposure time however is as
observational problems such as high altitude cirrus clouds or bad seeing
diminish the quality of certain images which were not used for creating the
final stacks. As seen from Tab. \ref{observations} the time efficiency of the
three bands is in the order of 65\%.

We should note here that the V-band observations were taken using both
the blue and the red arms of the telescope. The blue arm was used during SDT,
and the red during LBTB time. Although the response curves of the two V-band
filters are identical, the quantum efficiencies of the detectors are slightly
different. For the analysis presented in this paper this effect is not
significant and we merge the two ($V_b$ and $V_r$) images to achieve greater
depth. However in more detailed studies, one should treat the $V_b$ and $V_r$
images separately.

\begin{table}
\label{observations}
\centering
\begin{tabular}{llccc}
\hline\hline
                      &                  &  U   &  B  &  V  \\
\hline
\multirow{15}{*}{SDT} & 21 February 2007 &      & 60  &     \\
                      & 23 February 2007 &      & 30  &     \\
                      & 25 February 2007 &      & 60  &     \\
                      & 15 March 2007    &      &     & 54  \\
                      & 16 March 2007    &      &     & 114 \\
                      & 02 April 2007    &      & 30  &     \\
                      & 11 April 2007    & 30   & 30  &     \\
                      & 10 May 2007      & 60   &     &     \\
                      & 11 May 2007      & 54   &     &     \\
                      & 12 May 2007      & 54   &     &     \\
                      & 19 May 2007      & 90   &     &     \\
                      & 20 May 2007      & 60   &     &     \\
                      & 21 May 2007      & 36   &     &     \\
                      & 22 May 2007      & 60   &     &     \\
                      & 11 June 2007     & 60   &     &     \\
\hline
\multirow{9}{*}{LBTB} & 07 March 2008    & 90   & 120 & 120 \\
                      & 08 May 2008      & 168  &     &     \\
                      & 09 May 2008      & 78   &     &     \\
                      & 11 May 2008      & 240  &     &     \\
                      & 29 December 2008 &      & 60  & 60  \\
                      & 30 December 2008 & 60   & 54  & 30  \\
                      & 31 December 2008 & 63   &     &     \\
                      & 01 March 2009    &      & 69  & 69  \\
                      & 02 March 2009    & 51   &     &     \\
\hline
SDT                   &                  & 504  & 210 & 168 \\
LBTB                  &                  & 750  & 303 & 279 \\
Total                 &                  & 1254 & 513 & 441 \\
Effective             &                  & 828  & 333 & 321 \\
\hline\hline
\end{tabular}
\caption{Details of the observing runs (time is in minutes). The SDT
         runs were made using the blue channel only, whereas the LBTB runs used
         both telescopes. The effective exposure time is the time actually used
         for the final stacks.}
\end{table}


\section{Data reduction}

For the reduction of the data we have primarily used IRAF routines included
in the \emph{mscred} package, which is designed to reduce mosaic data.

\subsection{Initial calibration}

Initial corrections to remove the ``pedestal'' level of each chip have been
carried out using the overscan regions. We have found that the level of the
corrections varies significantly (up to 5\%) with the column of the chip and
therefore we fitted an 8-order Legendre polynomial to it. Residual
errors (possibly row-dependent) have been corrected using bias frames taken
at the beginning and end of each night with zero integration time.

Flat corrections have been made using sky flat frames taken at either dusk or
dawn (or both) for each filter at each arm of the telescope separately. A
master flat has been created for each night, filter, and arm. We divided
the bias-corrected images with their respective flat fields and noticed that
the outer edges of the fourth chip, corresponding to the outer edges of the
field of view were extremely noisy, possibly due to poor illumination. We
flagged them as bad regions.

After flat-fielding, we corrected the images for bad pixels. These include
columns of the CCD with non linear response or dust on the CCD surface. Bad
pixel masks are created and the correction has been made by interpolating
the values of neighboring pixels. Note that because of the absence of
neighboring pixels on the edges of the field of view, these areas are not
corrected and are simply not taken into account for the final stages when
the images are stacked.

Finally, there are some bright sources in the field which saturate the response
of the CCD. In the most severe cases the flux is so high that the current
affects the neighboring pixels, leaving ``bleeding trails''. This has a
significant effect to the B and V images and therefore these regions have
been identified and masked out.

\subsection{Astrometry}

\begin{figure}
  \resizebox{\hsize}{!}{\includegraphics{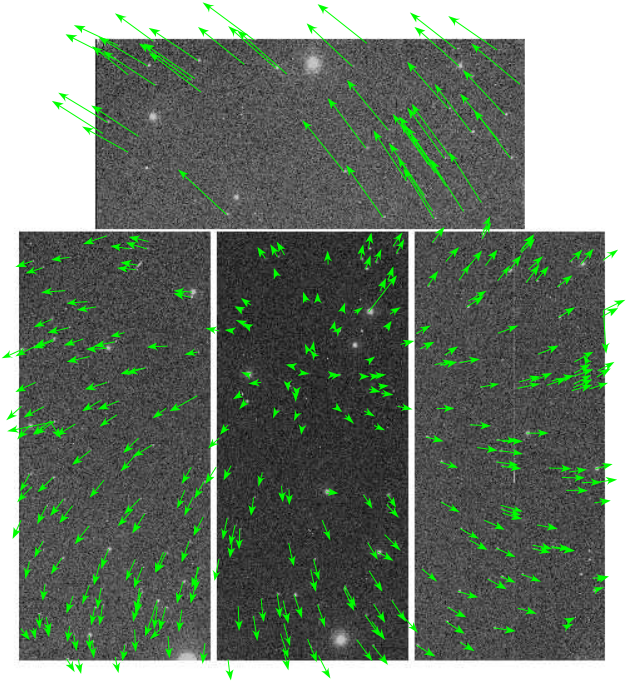}}
  \caption{Astrometrical solutions applied to an exposure image of 11 June 2007
           (U-band). The image has been bias and flat calibrated but no
            astrometrical solution has been applied. The arrows represent the
            shifts to the calibration sources exaggerated 8 times. To
            compensate for the distortion we dealt with each chip separately
            and after applying an initial shift we fitted a 4th order
            polynomial to the corrections. The corrected images have an rms
            scatter of $\sim$0.2\,arcsec}
  \label{astrometry}
\end{figure}

Before dealing with the (arcsecond-scale) astrometric errors, we correct each
image for an initial offset, in the order of several arseconds, caused by the
telescope's pointing inaccuracy. We use the brightest ($r<19$) sources from
the USNO-A2 \citep{Monet1998} catalogue to correct for this offset. This is
done by simply updating the wcs header of each file to match the coordinates
of the catalogue stars.

After having done that, we need to correct for the true astrometric errors
caused by the camera distortion. For this
purpose we do not use the USNO catalogue of the brightest stars, as proper
motions could have an effect in the solutions we derive. We use an astrometry
corrected catalogue of the Lockman Hole, which includes sources brighter than
$V=19$. This is based on observations made with the Canada-France-Hawaii
Telescope \citep[CFHT, e.g.][]{Wilson2001} and the data reduction details are
in \citet{Kaiser1999}. The absolute astrometry of this catalogue is based
on USNO-A2 which claimed accuracy is 0.25\,arcsec \citep{Monet1998}.

\begin{figure*}
  \resizebox{\hsize}{!}{\includegraphics{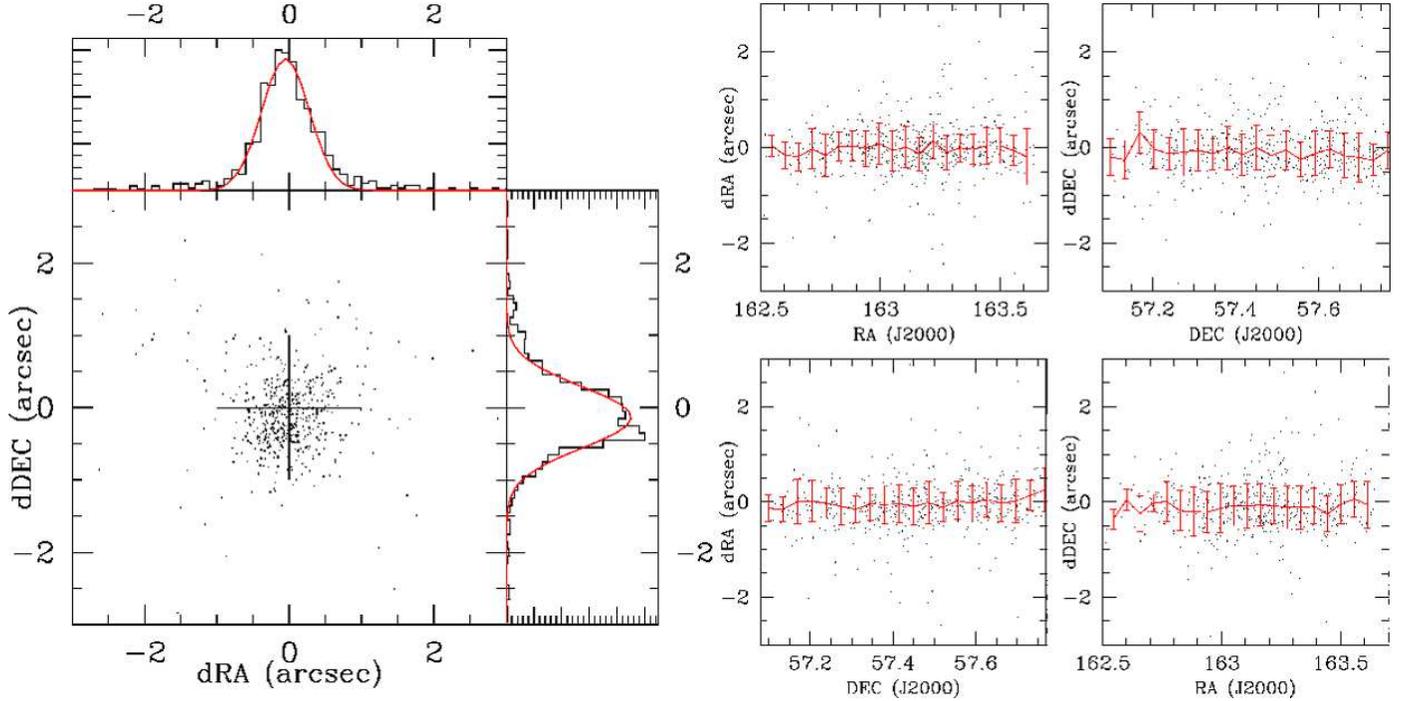}}
  \caption{Final astrometry check using the USNO-A2 catalogue. The left panel
           shows the RA and DEC differences between the LBT and USNO
           counterparts. The (gaussian) fits to the respective histograms
           have $\sigma=0.47$\,arcsec and $\sigma=0.59$\,arcsec for RA and DEC
           and the means are at -0.06\,arcsec and -0.13\,arcsec respectively.
           The right panels show the RA and DEC differences across the
           entire LBT image, where we do not observe any systematic trends.}
  \label{astrocheck}
\end{figure*}

To apply detailed astrometrical solutions we deal with each
chip separately in order to avoid fitting for jumps between the chips. We first
apply a distortion pattern which we empirically derived by correcting a random
image and then fit a 4-order polynomial to each direction of each chip. The
final rms scatter we get is in the order of ~0.2 arcsec. An example of the
astrometrical solutions applied (after correcting for the overall pointing
offset) is
given in Fig. \ref{astrometry}. To measure the final astrometrical accuracy,
we compare our LBT images with the USNO-A2 catalogue (see
Fig. \ref{astrocheck}) and with others, such as USNO-B1, APM \citep{Irwin1994},
SWIRE-IRAC(3.2$\mu$m) \citep{Lonsdale2003} and L-band VLA \citep{Biggs2006}.
Their positions typically agree within 0.4\,arcsec and the typical standard
deviation is 0.45\,arcsec, which is the value we assume to be our final
astrometric accuracy. The relative astrometry of the U-B-V images based on the
positions of bright ($<$24\,mag) and realtively compact (FWHM$<$1.5\,arcsec)
sources has a standard deviation of 0.066\,\arcsec.

\subsection{Background subtraction}

The final step is to subtract the sky background. After having flattened the
images and having corrected for field distortions (without preserving the
flux of each pixel) the sky background is uniform within a good approximation.
To subtract it we fragment the image to a grid constructed of 100x100 pixel
wide meshes and smooth each mesh by a median filter with a 5x5 pixel kernel
using sextractor \citep{Bertin1996}.
We chose this method over fitting a function to the
background because it gives better results in the vicinity of bright stars in
the sense that it does not over-correct the background.

\begin{figure*}
  \resizebox{\hsize}{!}{\includegraphics{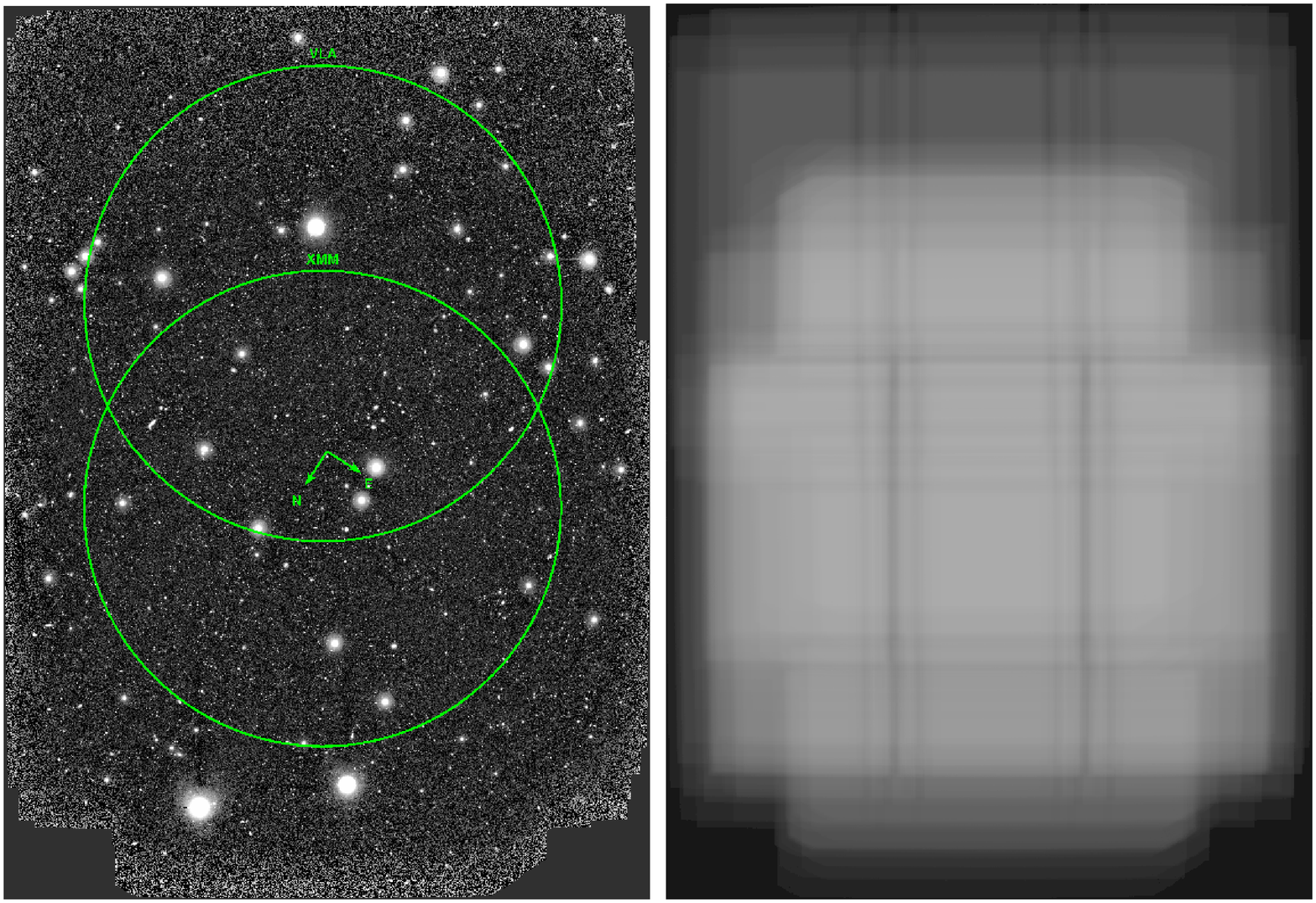}}
  \caption{Final U-band image and corresponding exposure map. The image covers
           both the XMM and the VLA fields, which are marked with
           10\arcmin-radii circular regios, close to their typical widths. The
           highest exposure is in the overlapping region between them.}
  \label{final_image}
\end{figure*}

After having subtracted the background, we re-project the four chips of each
image to a common frame, applying the complex astrometrical solutions. By
doing that we get rid of complex headers and multiple frame images. We use
the same reference image to re-project all images in all filters. Finally,
after removing any bad images due to poor observing conditions or other
problems, we stack all the images (weighted according to their exposure
times) to produce the U, B, and V maps of the Lockman Hole. An example image
(in the U band) and its corresponding exposure map is shown in Fig.
\ref{final_image}. The regions mark the deep VLA and XMM surveys with 10\arcmin
radii, which are their typical widths.

\subsection{Flux calibration}

For the B and V filters we rely on flux-calibrated images of the Lockman Hole
taken with the Calar Alto Telescope \citep[see][]{Kaiser1999,Wilson2001}. We
select point-like sources
which are not saturated in any of the images and conduct aperture photometry.
We compare the results and derive zero-point magnitudes for our final images.
We do not find evidence for a gradient across the image.

As there are no sources with known magnitudes in the U-spec band, we had to
rely on U-Bessel standards to derive the zero-point offsets. We used the
observations of June 11, 2007 when standard stars are observed with both
the U-Bessel and the U-spec filters. We have applied the same calibration
(bias subtraction and flat fielding) using the same bias and flat-field images
to all the target
and standard star frames and did not perform any astrometric corrections nor
we combined the calibrated files. We measured the observed magnitudes of four
standard stars without applying any zero-point offsets and found a difference
of $\sim0.67$ mag. We attribute this difference to the higher
efficiency of the U-spec filter, as the spectral profiles are similar.

We then calculate the zero-point offset for the U-Bessel filter using the
equation:
\[U_{Bessel}[zp]=U_{Bessel}-u_{Bessel}-k_{U}X-c.t.(U-B)\]
where $U_{Bessel}$ and $u_{Bessel}$ are the correct and observed magnitudes,
$k_{U}$ is the extinction term for the U-Bessel filter, $X$ the airmass of
the observation and c.t. the colour term. To have consistency between the
different observations made with different integration times, we have adopted
everything to $t_{int}=1$\,s. During the commissioning of the LBC-blue the
extinction term for the U-Bessel filter was measured to be
$k_{U}=-0.48\pm0.02$ and the colour term $c.t.=0.036$\footnote{The LBC
commissioning report can be found at
\texttt{http://lbc.oa-roma.inaf.it/commissioning/}}. Applying the
$U_{Bessel}$, $u_{Bessel}$, and $X$ values
of the standard stars we derive:
$U_{Bessel}[zp]=(26.012\pm0.014)-2.5\log\frac{t_{\rm exp}}{1{\rm sec}}$.

To calculate the U-spec zero-point offset we shift the value of the U-Bessel
offset by the mean measured magnitude difference of the standard stars. This
is the equivalent of assuming that the standard stars have the same magnitudes
in the U-spec and U-Bessel filters. The central wavelengths and widths of
the two filters are very similar, so such an assumption does not affect the
result in great extent. We find: $U_{spec}[zp]=26.073\pm0.012$.

To calculate the zero-point offset of the final image, where as a result of
rescaling of the individual images and combining them the connection to the
original gains has been lost, we use the number mentioned in the previous
paragraph to derive the magnitudes of Lockman Hole sources using the raw
images. For this purpose we selected 43 non
saturated sources with almost gaussian profiles which are observed with the
second chip of the mosaic, as the standard stars. We use these 43 sources to
measure the zero-point magnitude of the final combined image. We derive
$zp_{final}=32.041\pm0.013$ and do not find any evidence for a gradient in any
direction of the image. The magnitudes of the standard stars are given in the
Landolt photometric system \citep{Landolt1992} which is based on Vega
magnitudes. We use U-Bessel AB correction calculated during the commissioning
time (0.87), so our final zero-point offset for the U band is:
$U_{spec,AB}[zp]=32.911\pm0.013$.

Detailed information on the three (U-B-V) final images can be found in Tab.
\ref{image_info}.

\begin{table}
\centering
\begin{tabular}{cccccc}
\hline\hline
  & zero-point &  seeing  &   number   & 50\% eff. & 4.5\,$\sigma$ \\
  &  (AB mag)  & (arcsec) & of sources & (AB mag)  &   (AB mag)    \\
\hline
U &   32.911   &   1.06   &   51500    &   26.7    &     28.9      \\
B &   33.830   &   0.94   &   76071    &   26.3    &     29.1      \\
V &   34.110   &   1.03   &   68278    &   26.3    &     28.6      \\
\hline\hline
\end{tabular}
\caption{Flux calibration, quality and source extraction information of the
         images.}
\label{image_info}
\end{table}


\section{Results}

\subsection{Source catalogues}

The source detection has been done independently in each of the U-B-V images
using sextractor. Sources are identified as regions where 12 or more adjacent
pixels have values above 1.2 times the local background rms. The algorithm
first subtracts the background which is fitted by segmenting the image with a
grid. If the grid is too fine, a fraction of the flux of the sources will be
subtracted as background and this will be more severe for extended sources.
On the other hand, a very large mesh will fail to subtract the background
near very bright objects, where stray light contaminates the image, so the
source extraction will fail in these areas. To overcome these issues we ran
sextractor in two steps: first we used a very fine grid (with a $5\times5$
pixel mesh) to subtract the background and created a ``source detection''
image. We then re-run sextractor in dual mode using this image to detect
the sources but measure their fluxes from the original image, where the
background is subtracted using a $10\times10$ pixel mesh.
\begin{figure}
  \resizebox{\hsize}{!}{\includegraphics{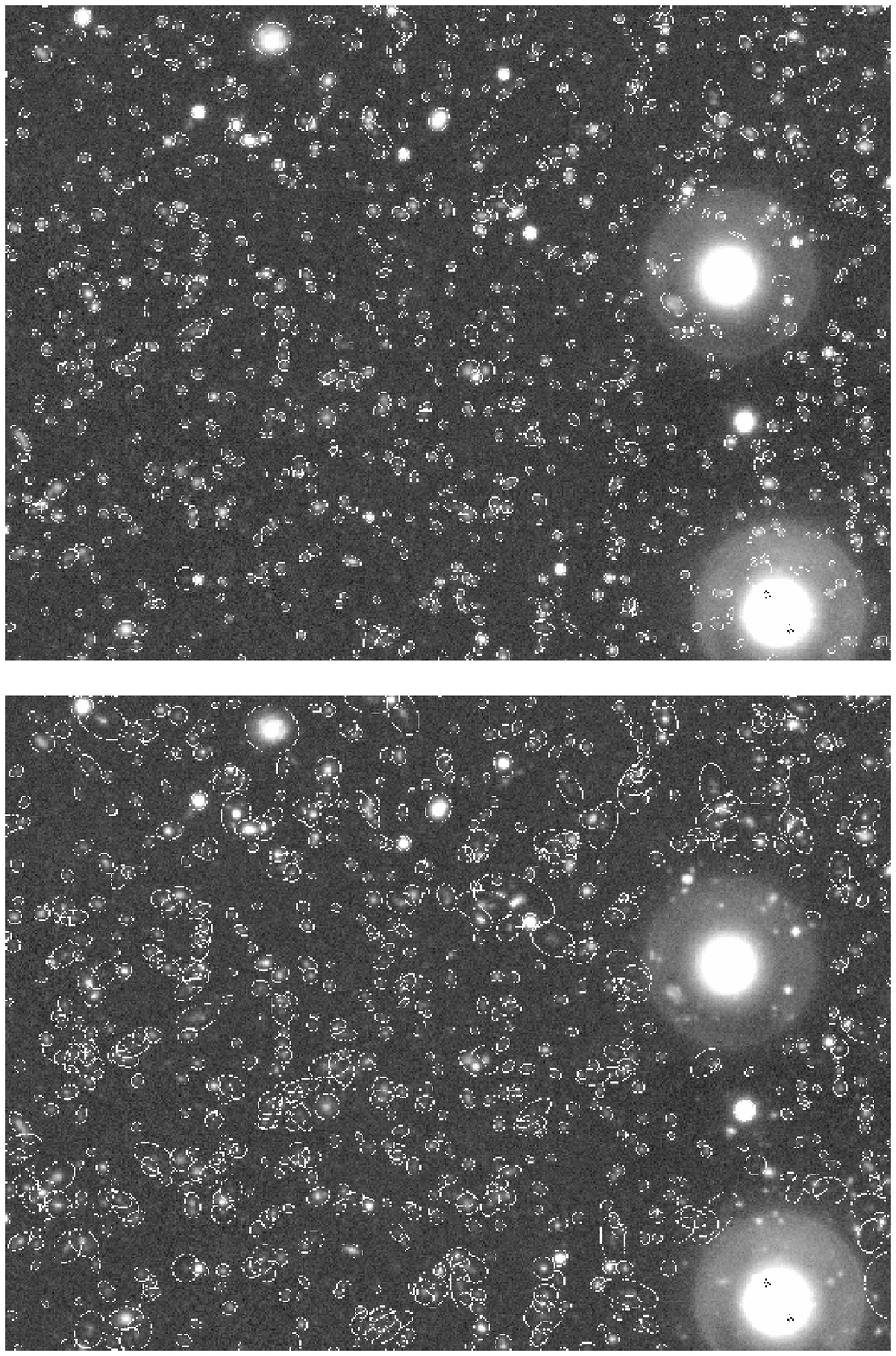}}
  \caption{Apertures of extracted sources in the U image. The upper panel
           shows the ``dual mode'' results and the lower the ``single mode''
           (see text) Note that the detection efficiency near the bright
           object as, the deblending efficiency and the apertures of faint
           sources are much better in ``dual mode'', but the apertures of
           bright objects are more reasonable in ``single mode''.}
  \label{extraction}
\end{figure}
This method has the drawback that the apertures where we measure the flux
are too tight for bright sources, which appear more extended in the images
and as a consequence we are losing a fraction of their flux (see Fig.
\ref{extraction}, upper panel). Therefore we run sextractor once more in
``single mode'' (with a $10\times10$ pixel mesh for the background; see also
Fig. \ref{extraction}, lower panel) and replace the sources with magnitudes
brighter than 22 of the original catalogue with those extracted in ``single
mode''.

Finally, in order to avoid spurious detections, we remove from our catalogues
sources whose isophotal flux errors are larger than the fluxes and therefore
do not have reliable photometry and sources whose FWHM is less than 90\% of
the seeing of each image (1.06, 0.94, and 1.03\,arcsec for the U, B, and V
images respectively) and are related with imaging artifacts. We also
optically inspect the remaining sources and remove obvious false detections
related with bad pixels, dust on the CCDs, bleeding trails etc as well as
saturated sources. The final U, B, and V catalogues contain 51500, 76071,
and 68278 sources respectively.

\begin{figure*}
  \resizebox{\hsize}{!}{\includegraphics{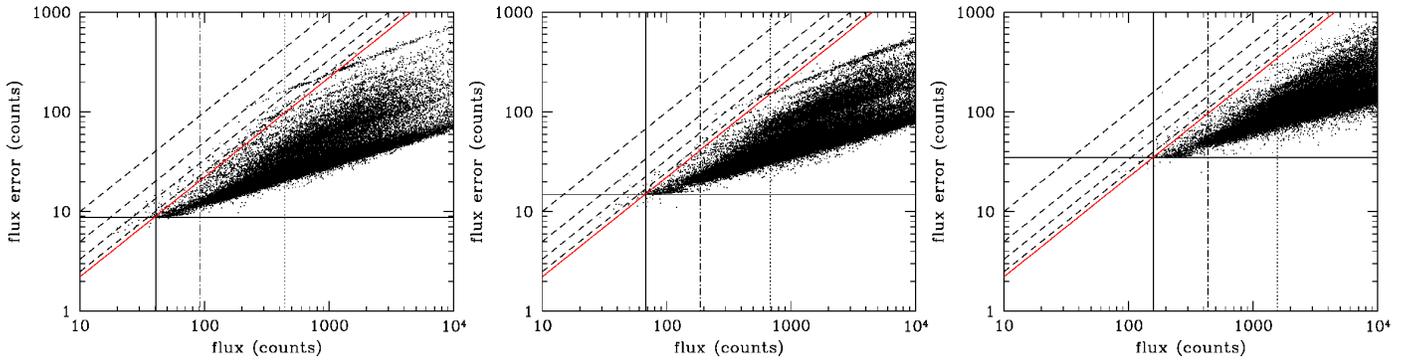}}
  \caption{Flux error versus flux diagrams for source detections in the U, B,
           and V filters. The zero-point magnitudes for the three filters are
           $32.911\pm0.013$, $33.830\pm0.014$, and $34.110\pm0.011$
           respectively.
           The dashed lines represent signal-to-noise ratios of 1, 2, and 3,
           and the red solid lines represent the (empirical) limit of the
           source extracting algorithm used ($4.5\,\sigma$). The horizontal
           and vertical solid lines mark the minimum flux and the respective
           minimum flux error of the faintest sources extracted. The dotted
           and dash-dotted lines mark the 50\% efficiency limit and the limit
           of the number count analysis respectively.}
  \label{limits}
\end{figure*}

In order to estimate the detection limits of our catalogues we plot the flux
error against the flux of each detected source (see Fig. \ref{limits}). We
do this because we used a more complex selection algorithm to extract sources
than a simple signal-to-noise cut. The dashed lines in Fig. \ref{limits}
represent signal-to-noise ratios of 1, 2, and 3 from left to right and the
red line the (empirical) ``detection limit''. We notice that the faintest
sources tend to be closer to this limit and this is a result of them being
point-like. The resulting detection threshold magnitudes and signal-to-noise
ratios (see Tab. \ref{image_info}) are not detection limits in the sense that
sources with fluxes (or SNRs) above these limits are detected, they are
indicative of the sensitivity of the survey representing the lowest flux
(and respective SNR) of securely detected sources. They also provide no
information on the completeness of the survey at a given flux (or SNR), nor
an estimation of the chance of a spurious detection. Such an analysis is
described in \S \ref{number_counts}. 

\subsection{Colours}

To create colour catalogues of the various sources detected in the U, B, and
V images, one could simply cross-correlate the three source catalogues
described in the previous paragraphs and compare the fluxes in the different
bands. This however would introduce an uncertainty on the choice of the best
counterpart and moreover the deblending efficiency of sextractor varies
between the different images, so a source in one catalogue might be blended
with a close pair in another or vice-versa. Therefore we chose to select
one image to extract the sources and then measure their fluxes using the
other images in dual mode.

We make the source detection on a combined image of the three bands, the
$\chi^2$-image. The PSFs of the three images we combine do not have significant
differences; the worse PSF (U-band) is only 6\% larger than the best (B-band),
therefore we do not lose in quality when combining the images as compared to
using the best PSF image and we gain in S/N. We follow the recipe of
\citet*{Szalay1999} to create the $\chi^2$-image: after carefully removing any
residual background of each image (using the ``-BACKGROUND'' checkimage option
of sextractor) we fit the off-source pixel histogram with a gaussian, checking
that the noise profile is indeed gaussian. We then scale the three images
according to their noise amplitudes and we create the $\chi^2$-image, which is
the square root of the sum of the squares of the individual pixel values.
We then extract the sources from the combined image using the method described
in the previous paragraph and measure the fluxes in the individual U-B-V
images. Again, we consider a detection real if its FWHM is $>90\%$ of the
PSF FWHM of the best individual image (0.85\,arcsec). The final colour
catalogue contains 88429 with detections in all three bands.

\begin{figure}
  \resizebox{\hsize}{!}{\includegraphics{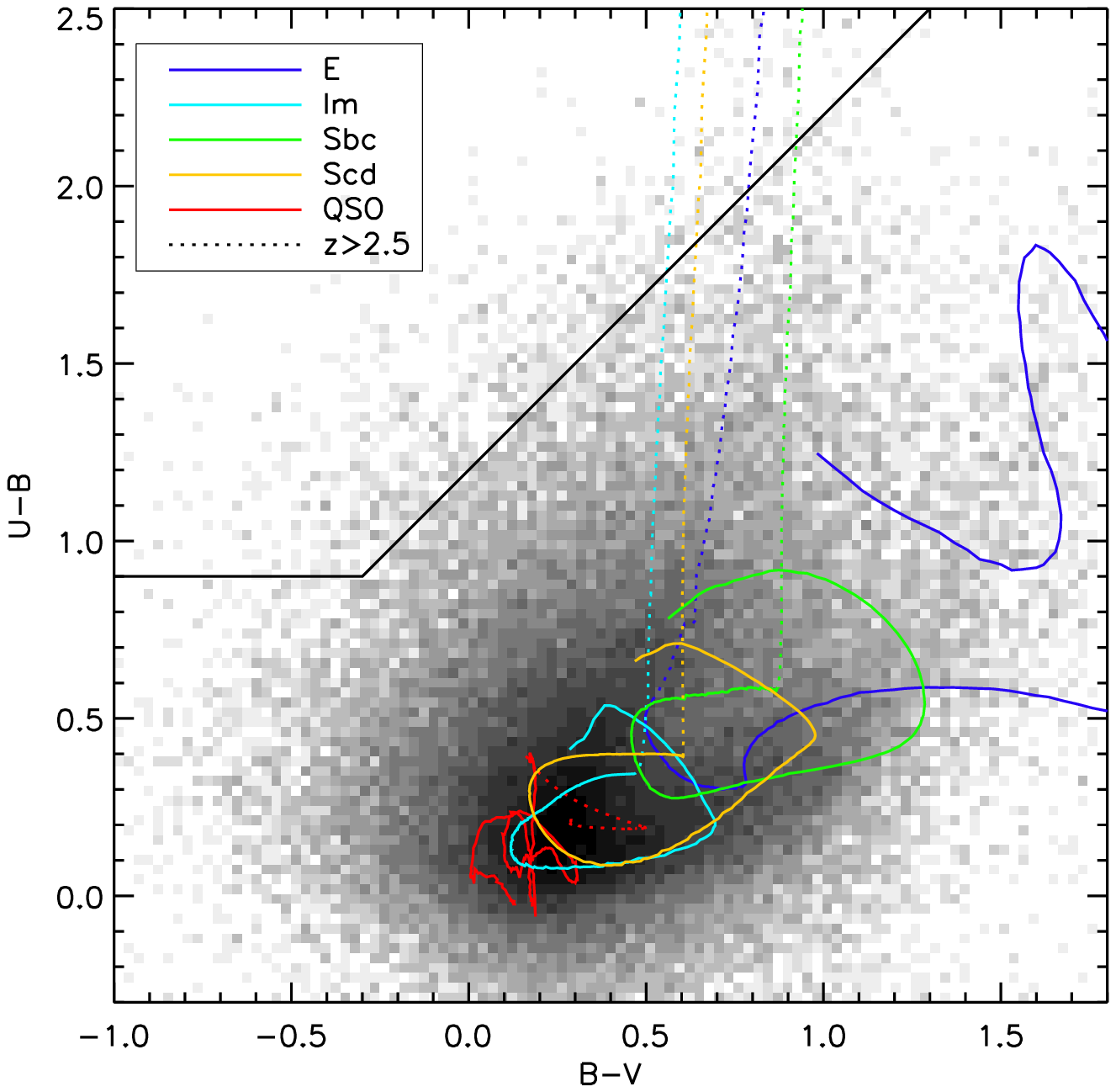}}
  \caption{Optical colours of the Lockman hole sources in the AB system.
           THe greyscale represents the density of sources detected in all
           three bands are plotted. The black line marks the selection area
           of U-dropout sources with redshift $z\gtrsim3$
           (see Fig.\,\ref{colours_z} and \S\,\ref{udrops}). Dotted are the
           tracks above $z=2.5$, where the SEDs are not well sampled.
           The coloured lines represent tracks of different
           SED templates: blue line for ellipticals, cyan for irregular,
           green for Sbc, yellow for Scd, and red for QSO.}
  \label{colours}
\end{figure}

A colour-colour ($U-B$\,vs.\,$B-V$) diagram of the Lockman Hole sources is
shown in Fig. \ref{colours}. The greyscale represents the density of sources
detected in all three bands and the black lines mark the selection area of
$z\gtrsim3$ objects. We also calculate the colours of different
galaxy SED templates from \citet*{Coleman1980} and a QSO template from
\citet{Cristiani1990}. The galaxy templates are extrapolated to the Lyman
break (911.25\,\AA) and are zeroed thereafter. The Lyman break meets the blue
end of the U filter at $z=2.5$ and this is the highest redshift where the
colour tracks are reliable (solid lines). The dotted lines $(z>2.5)$ are
shown as an approximation of the colours of high redshift galaxies.

\subsection{Number counts}\label{number_counts}

In order to derive the differential number counts of extragalactic
sources in the U, B, and V bands, we select a region in the centre of
the field with uniform exposure within a good approximation. This region
has a size of $14\times13.3$\,arcmin and is located in the area where the
XMM and VLA observations overlap.

The first step in calculating number counts is to estimate the
source extracting efficiency at a given magnitude. The way to do it is
to create an image with artificial sources of known magnitudes and to apply
the same source extracting procedure as applied to the image, and measure
the fraction of the sources recovered. We use the artdata package in IRAF
to create lists of artificial sources. They contain sources with a
uniform spatial distribution and magnitudes ranging from 16 to 29 following a
power-law distribution with a power of 0.5. The surface brightness profiles
are exponential discs (resembling spirals) and $r^{-1/4}$ discs, resembling
ellipticals. The fraction of elliptical galaxies in the random catalogues is
20\% \citep[see][]{vandenBergh2001}. Here, we caution that adding a large
number of artificial sources in the image might change its crowding properties,
however we need a large sample of sources for reliable statistics. To avoid
confusion, we create a list of 100000 sources and split it to 100 1000-source
samples.

\begin{figure*}
  \resizebox{\hsize}{!}{\includegraphics{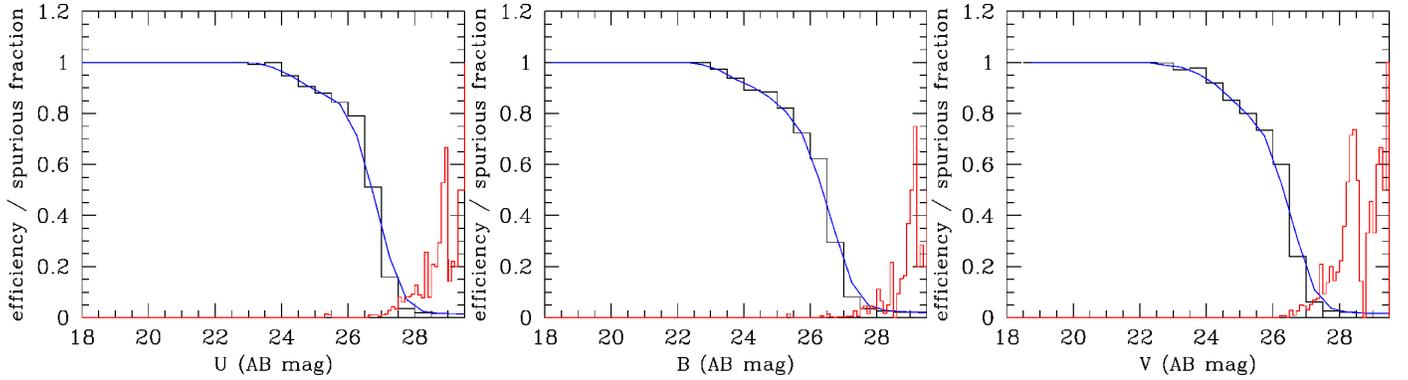}}
  \caption{Detection efficiency histograms of the source detection
           algorithm used, derived from simulations. The blue lines
           represent the smoothed efficiencies used to correct the
           surface density distributions. The red histograms are the
           estimates of the fraction of detections which are spurious,
           using the artificial noise images.}
  \label{efficiency}
\end{figure*}

We plant these sources into the cutouts of the final images and apply the
same source extracting algorithm we used to create the source catalogues.
We then measure the fraction of the artificial catalogue we retrieve,
hence the efficiency of the source detecting method at any given magnitude
and average the results of the 100 subsamples. Increasing the number of
sources of each subsample we get similar results up to the point where the
number of sources is comparable to the number of ``real'' sources in the
region ($\sim10000$).
The results for all three bands are shown in Fig. \ref{efficiency}.
This method has the drawback that the artificial sources are mixed with real
sources, and so there is no way of knowing whether a detected source is real
or an artifact. The surface density of sources with magnitude (at any band)
$<27$ is close to $2\times10^{5}\rm{deg}^{-2}$, which means that there is a
$\sim10$\% probability that a real source is within 1.5\,arcsec of a random
position.

To measure the spurious source detection rate we measure the off-source noise
of the science images and check that the noise profile is gaussian. We then
create gaussian noise maps of the same amplitude and insert the artificial
sources there. In this case it is desirable to reproduce the crowding of the
original field, so we include a large number of artificial sources (30000),
which is the number of sources with magnitude $<29$ we expect in this field.
The source extraction output to these composite images provides
the information of the spurious detection rate, plotted with the red lines
in Fig. \ref{efficiency}. We can see that the number of spurious sources is
negligible below 26.5\,mag and starts becoming important above 28.0\,mag,
where the efficiency drops to practically unusable values. From these
diagrams we can also derive the magnitude where the detecting efficiency
drops below 0.5, which is a meaningful measure of the detection threshold of
the image. This threshold is 26.7\,mag(AB), 26.3\,mag(AB), and 26.3\,mag(AB)
for the U, B, and V bands respectively.

\begin{figure}
  \resizebox{\hsize}{!}{\includegraphics{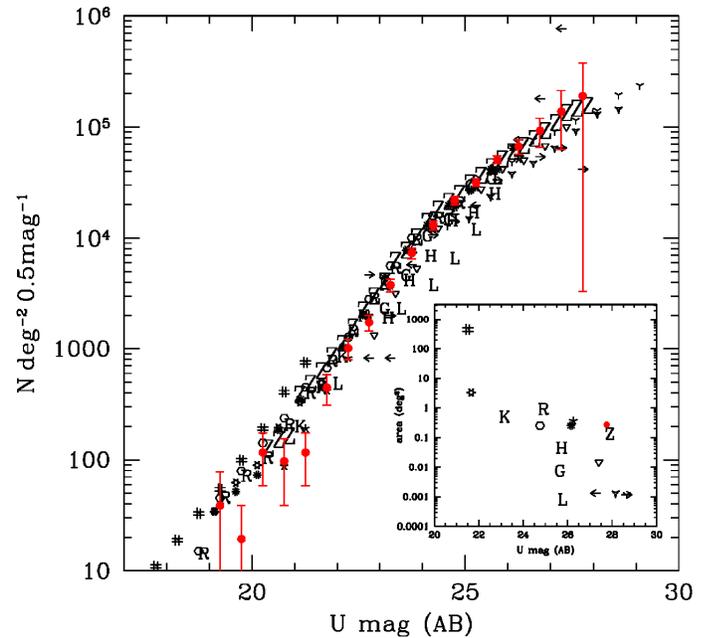}}
  \caption{U-band surface density distribution corrected for efficiency
           and spurious detections. The data points of this study
           (seen in Tab. \ref{numbercounts}) are plotted in red
           dots, while various symbols represent the results of other                      surveys found in the literature (see Tab. \ref{symbols}), corrected to AB magnitudes.
           The inner plot shows the depth of each survey (in terms of secure
           number count measurements) with respect to its width.}
  \label{Ucounts}
\end{figure}
\begin{figure}
  \resizebox{\hsize}{!}{\includegraphics{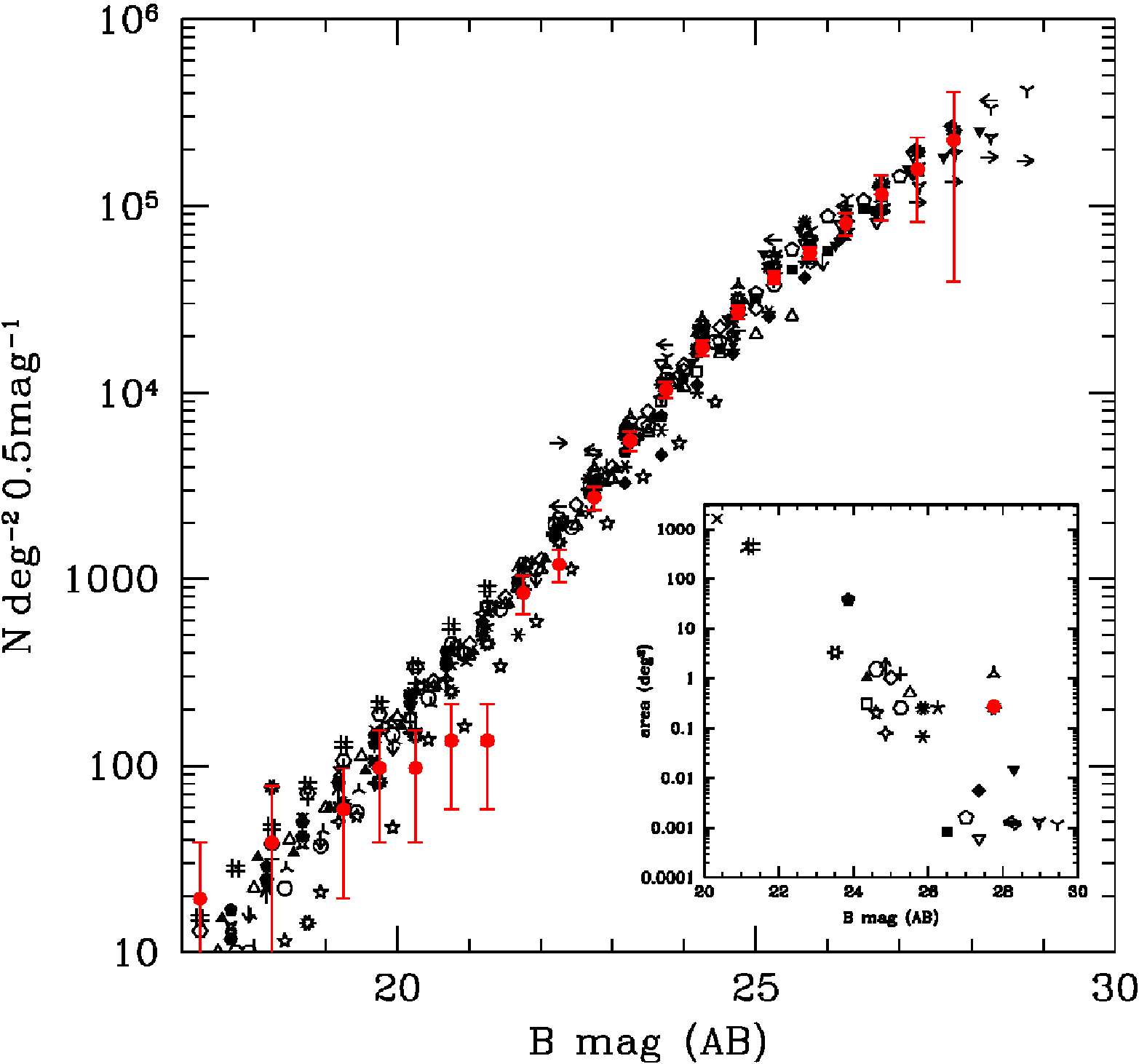}}
  \caption{Same as Fig. \ref{Ucounts} but for the B-band}
  \label{Bcounts}
\end{figure}
\begin{figure}
  \resizebox{\hsize}{!}{\includegraphics{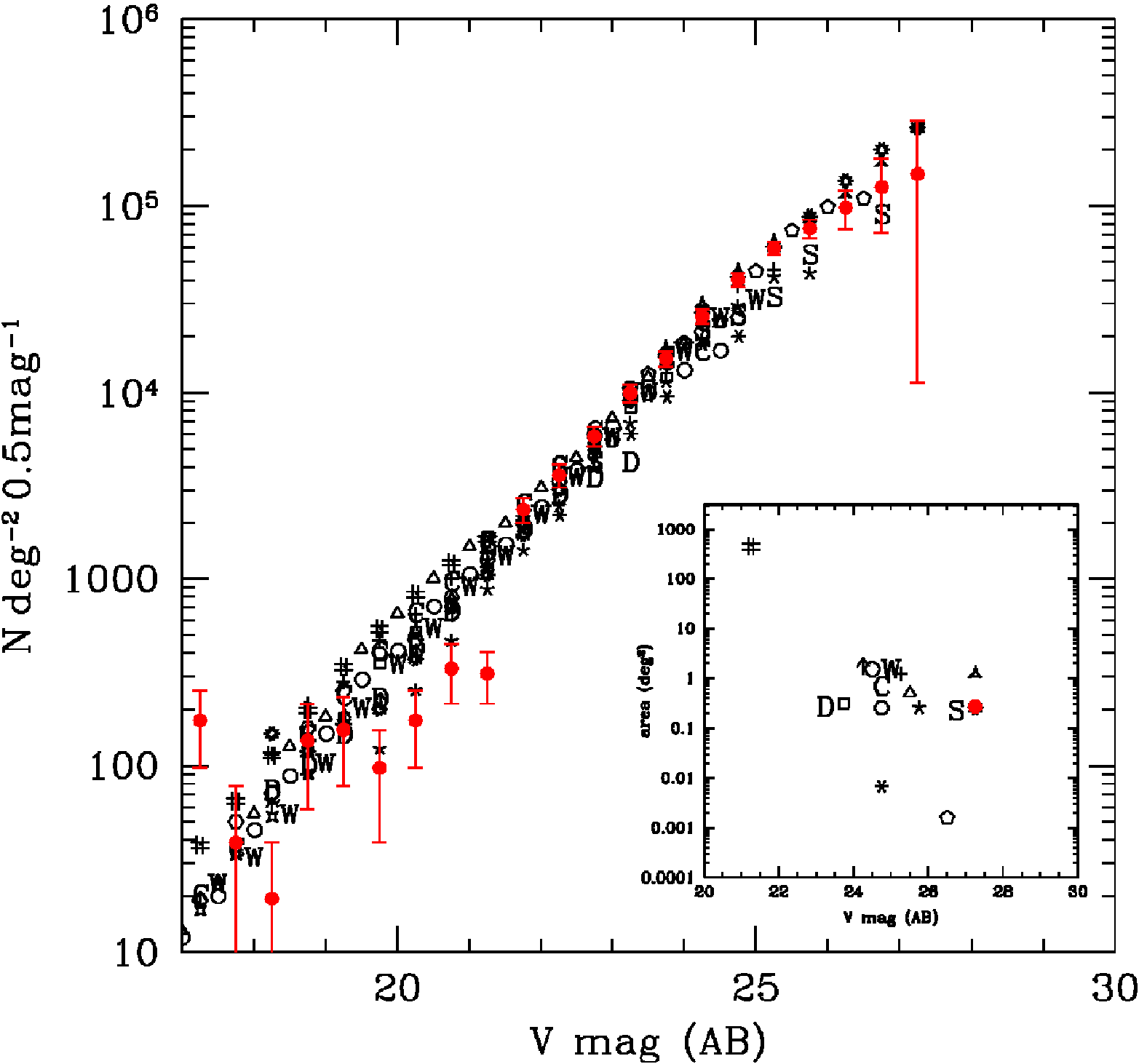}}
  \caption{Same as Fig. \ref{Ucounts} but for the V-band}
  \label{Vcounts}
\end{figure}

Figures \ref{Ucounts}-\ref{Vcounts} show the surface density distributions
for the three bands observed. Data points of other studies
found in the literature are also plotted. We have binned the
magnitudes of the observed sources in bins of 0.5\,mag. For each bin
we corrected the source counts using the efficiency and spurious
detection information. As we are interested in galaxy counts, we need
a selection mechanism for stellar sources, and as such we use the
``stellarity index'' of sextractor. This estimate works well for bright
sources, but because fainter galaxies can appear point-like it fails
for larger magnitudes. As a limiting magnitude we choose 21.5(AB).
Below this limit the stellar
counts are anyway negligible with respect to the number of galaxies
\citep*[see e.g.][]{Jarrett1994}.
The error bars take into account Poisson
uncertainties of the uncorrected counts, efficiency uncertainties and
cosmic variance. For the latter we use the computational tool of
\citet{Trenti2008}, which compares the two-point correlation function of
dark matter with the volume of the survey. As a typical redshift for our
survey we use $z=1$ (see also \S\ref{nc_discussion}), though using a different
value in the range $0.5-2.5$ does not change the result in great extent. We
find that the cosmic variance uncertainty is important for bright magnitudes
(typically $<25.5$) where the number of intrinsic objects is relatively small.
For fainter magnitudes the efficiency uncertainties dominate. As an estimation
of those we choose the mean efficiency difference between the bin in question
and its neighboring bins. This way we account for the effects of binning,
in other words the different efficiencies the magnitudes within each bin
have.

\begin{table*}
\centering
\begin{tabular}{cccc}
\hline\hline
Reference                            & Symbol                  & Telescope         & Instrument                  \\
\hline
\citet{Alcala2004}                   & $\triangle$             & ESO/MPG 2.2\,m    & WFI                         \\
\multirow{2}{*}{\citet{Arnouts1997}} & \multirow{2}{*}{$\Box$} & ESO 3.60\,m       & EFOSC                       \\
                                     &                         & ESO-NTT 3.5\,m    & EMMI                        \\
\citet{Arnouts1999}                  & $\pentagon$             & ESO-NTT 3.5\,m    & SUSI                        \\
\citet{Arnouts2001}                  & $\hexagon$              & ESO/MPG 2.2\,m    & WFI                         \\
\citet{Berta2006}                    & $\Circle$               & ESO/MPG 2.2\,m    & ESIS                        \\
\multirow{3}{*}{\citet{Bertin1997}}  & \multirow{3}{*}{$\Yup$} & CERGA 0.9\,m      & \multirow{3}{*}{d-MAMA}     \\
                                     &                         & ESO               &                             \\
                                     &                         & SERC              &                             \\
\citet{Cabanac2000}                  & C                       & CFHT              & UH8K                        \\
\citet{Capak2004}                    & $\APLstar$              & Subaru            & Suprime                     \\
\citet{Driver1994}                   & $\hexstar$              & WHT               & Hitchhiker                  \\
\citet{Drory2001}                    & D                       & Calar Alto 2.2\,m & CAFOS                       \\
\citet{Eliche-Moral2006}             & \ding{89}               & INT La Palma      & WFC                         \\
\citet{Heydon-Dumbleton1989}         & $\times$                & UKST              & d-COSMOS                    \\
\citet{Furusawa2008}                 & \mytristar              & Subaru            & Suprime                     \\
\citet{Gardner1996}                  & \rotatebox[origin=c]{45}{\ding{71}} & KPNO 0.9\,m & T2KA                  \\
\citet{Grazian2009}                  & Z                       & LBT               & LBC                         \\
\citet{Guhathakurta1990}             & G                       & CTIAO             & pr. focus CCD               \\
\citet{Hogg1997}                     & H                       & Hale Telescope    & COSMIC                      \\
\multirow{2}{*}{\citet{Huang2001}} & \multirow{2}{*}{\ding{73}} & Calar Alto 2.2\,m & \multirow{2}{*}{CCD camera} \\
                                     &                         & Calar Alto 2.5\,m &                             \\
\citet{Jones1991}                    & \myhexagon              & AAT               & pr. focus d-COSMOS          \\
\citet{Kashikawa2004}                & \mytenstar              & Subaru            & Suprime                     \\
\citet{Koo1986}                      & K                       & KPNO 4\,m         & photographic plates         \\
\citet{Kummel2001}                   & $\blacktriangle$        & Calar Alto 3.5\,m & Cassegrain CCD              \\
\multirow{2}{*}{\citet{Lilly1991}}   & \multirow{2}{*}{$\blacksquare$} & CFHT      & \multirow{2}{*}{NSF1-TI}    \\
                                     &                         & UH 2.2\,m         &                             \\
\citet{Liske2003}                    & \gplPt{PentF}           & INT La Palma      & WFC                         \\
\citet{Maddox1990}                   & \myfillhex              & UKST              & d-APM                       \\
\citet{McCracken2001}                & $\Diamond$              & CFHT              & UH8K                        \\
\citet{McCracken2003}                & $+$                     & CFHT              & CFH12K                      \\
\citet{Metcalfe1991}                 & \ding{71}               & INT La Palma      & RCA (prime focus)           \\
\citet{Metcalfe1995} (a)             & $\Diamondblack$         & INT La Palma      & RCA (prime focus)           \\
\citet{Metcalfe1995} (b)             & $\triangledown$         & WHT La Palma      & Tek CCD (aux Cass)          \\
\citet{Metcalfe2001} (a)             & $\Ydown$                & HST               & WFPC2                       \\
\citet{Metcalfe2001} (b)             & \mytrjstar              & HST               & WFPC2                       \\
\citet{Metcalfe2001} (c)             & $\blacktriangledown$    & WHT               & Tek CCD (pr. focus)         \\
\citet{Prandoni1999}                 & $\uparrow$              & ESO-NTT 3.5\,m    & EMMI                        \\
\citet{Radovich2004}                 & R                       & ESO/MPG 2.2\,m    & WFI                         \\
\citet{Smail1995}                    & S                       & Keck              & LRIS                        \\
\multirow{2}{*}{\citet{Songaila1990}} & \multirow{2}{*}{L}     & CFHT              & \multirow{2}{*}{NSF1-TI}    \\
                                     &                         & UH 2.2\,m         &                             \\
\citet{Tyson1988}                    & $\downarrow$            & CTIO              & prime focus CCD             \\
\citet{Volonteri2000}                & $\leftarrow$            & HST               & WFPC2                       \\
\citet{Williams1996}                 & $\rightarrow$           & HST               & WFPC2                       \\
\citet{Wilson2003}                   & W                       & CFHT              & UH8K                        \\
\citet{Yasuda2001}                   & \#                      & SDSS telescope    & SDSS imager                 \\
\hline\hline
\end{tabular}
\caption{Number counts data found in the literature. The symbols noted are
         used in Figures \ref{Ucounts}-\ref{Vcounts}. A ``d-'' prefix before
         the instrument symbolizes digitization of photographic plates.}
\label{symbols}
\end{table*}

\begin{table*}
\centering
\begin{tabular}{crrcrrcrrc}
\hline\hline
mag       &      U & $\sigma_U$ & eff   &      B & $\sigma_B$ & eff   &     V  & $\sigma_V$ & eff   \\
(AB)      & \multicolumn{2}{c}{(N\,deg$^{-2}$\,$(0.5\,mag)^{-1}$)} & & \multicolumn{2}{c}{(N\,deg$^{-2}$\,$(0.5\,mag)^{-1}$)} & &  \multicolumn{2}{c}{(N\,deg$^{-2}$\,$(0.5\,mag)^{-1}$)} &        \\
\hline
17.0-17.5 &      - &      -     & 1.000 &     19 &     19     & 1.000 &    174 &         77 & 1.000 \\
17.5-18.0 &      - &      -     & 1.000 &      - &      -     & 1.000 &     39 &         39 & 1.000 \\
18.0-18.5 &      - &      -     & 1.000 &     39 &     39     & 1.000 &     19 &         19 & 1.000 \\
18.5-19.0 &      - &      -     & 1.000 &      - &      -     & 1.000 &    135 &         77 & 1.000 \\
19.0-19.5 &     39 &     39     & 1.000 &     58 &     39     & 1.000 &    155 &         77 & 1.000 \\
19.5-20.0 &     19 &     19     & 1.000 &     97 &     58     & 1.000 &     97 &         58 & 1.000 \\
20.0-20.5 &    116 &     58     & 1.000 &     97 &     58     & 1.000 &    174 &         77 & 1.000 \\
20.5-21.0 &     97 &     58     & 1.000 &    135 &     77     & 1.000 &    329 &        116 & 1.000 \\
21.0-21.5 &    116 &     58     & 1.000 &    135 &     77     & 1.000 &    309 &         97 & 1.000 \\
21.5-22.0 &    445 &    135     & 1.000 &    836 &    195     & 0.994 &   2359 &        367 & 1.000 \\
22.0-22.5 &   1005 &    213     & 1.000 &   1186 &    237     & 0.978 &   3600 &        502 & 0.999 \\
22.5-23.0 &   1726 &    291     & 0.997 &   2748 &    395     & 0.978 &   5826 &        729 & 0.989 \\
23.0-23.5 &   3762 &    504     & 0.997 &   5515 &    672     & 0.950 &   9933 &       1100 & 0.981 \\
23.5-24.0 &   7307 &    810     & 0.979 &  10336 &   1071     & 0.939 &  15204 &       1499 & 0.955 \\
24.0-24.5 &  12984 &   1282     & 0.950 &  17327 &   1609     & 0.925 &  25504 &       2295 & 0.915 \\
24.5-25.0 &  21650 &   1933     & 0.910 &  27021 &   2317     & 0.918 &  40249 &       3340 & 0.856 \\
25.0-25.5 &  31629 &   2671     & 0.876 &  41229 &   3298     & 0.891 &  59463 &       4564 & 0.794 \\
25.5-26.0 &  50749 &   3950     & 0.837 &  55918 &   4307     & 0.844 &  75514 &       8690 & 0.711 \\
26.0-26.5 &  66532 &   9903     & 0.714 &  80376 &  11135     & 0.757 &  97480 &      22917 & 0.525 \\
26.5-27.0 &  92452 &  27568     & 0.486 & 114551 &  31570     & 0.584 & 125436 &      54593 & 0.301 \\
27.0-27.5 & 137413 &  75439     & 0.235 & 156755 &  75783     & 0.357 & 147554 &     136349 & 0.109 \\
27.5-28.0 & 189647 & 186364     & 0.071 & 223749 & 184589     & 0.155 &      - &          - &     - \\
\hline\hline
\end{tabular}
\caption{Differential number counts (in 0.5\,mag bins), uncertainties and
         source detections efficiencies for each magnitude bin of galaxies in
         the U, B, and V bands}
\label{numbercounts}
\end{table*}

The surface density data can be seen in Tab. \ref{numbercounts}. We
compare them with the results of various other studies (Figures \ref{Ucounts},
\ref{Bcounts}, \ref{Vcounts}, Tab. \ref{symbols}) and we are in good agreement. The small
diagrams of Figures \ref{Ucounts}, \ref{Bcounts}, and \ref{Vcounts} plot
the depth reached by each survey presented with respect to its covered area
(the various surveys used to create these figures are presented in Table
\ref{symbols}). The LBT survey presented here is the deepest one in the U
band ever conducted in such a large area and among the deepest in the B and
V bands. Significantly lower limits have been achieved only with the Hubble
Space Telescope in pencil-beam surveys (HDF-N and HDF-S), and these are highly
sensitive to cosmic variance \citep[see][]{Sommerville2004}. Comparing or
results with those from surveys of similar widths (made with the LBT and the
Subaru telescope) we find very good agreement.


\section{Discussion}

\subsection{Number counts}
\label{nc_discussion}

The models compiled by \citet{Metcalfe1996} and \citet{Metcalfe2001}
(normalized to 18\,mag using all data available in the bibliography) are
plotted against our measurements for the U and B bands in Fig.
\ref{number_fits_1}. We plot here three of the models presented in
\citet{Metcalfe1996} and \citet{Metcalfe2001}.
The short-dashed lines represent the pure luminosity evolution model, the
long-dashed line the same model with the inclusion of a population of
star-forming dwarf galaxies, which is the best-fit model in
\citet{Metcalfe1996} and the solid line is the same pure luminosity evolution
model with a modification of the faint-end slope of the luminosity functions
of late-type spirals ($\alpha=-1.75$ instead of $\alpha=-1.5$), used to fit
the multi-colour data of \citet{Metcalfe2001}. We find very good agreement
with the $\alpha=-1.75$ model in the B-band, although the U-band counts are
under-predicted by all models. However, the faint-end slope of the U-band
counts does seem to support a steepening of the faint-end slope of the LF.
\citet{Barro2009} have shown that the slope of the number count distribution
assymptotically reaches $-0.4(\alpha+1)$, where $\alpha$ is the faint-end
slope of the luminosty function if parametrized by a Schechter function.
Measuring the slopes of the number counts using the five faintest points
of each distribution, we calculate the faint-end slopes of the respective
luminosity functions: $\alpha_U=-1.733\pm0.018$, $\alpha_B=-1.748\pm0.006$,
$\alpha_V=-1.507\pm0.018$. We note that the assumed steep LF faint end slope
is in good agreement with the number count distributions of the U and B bands,
whereas the V band points to a LF with $\alpha=-1.5$
\citep[see][]{Metcalfe2001}.

\begin{figure}
  \resizebox{\hsize}{!}{\includegraphics{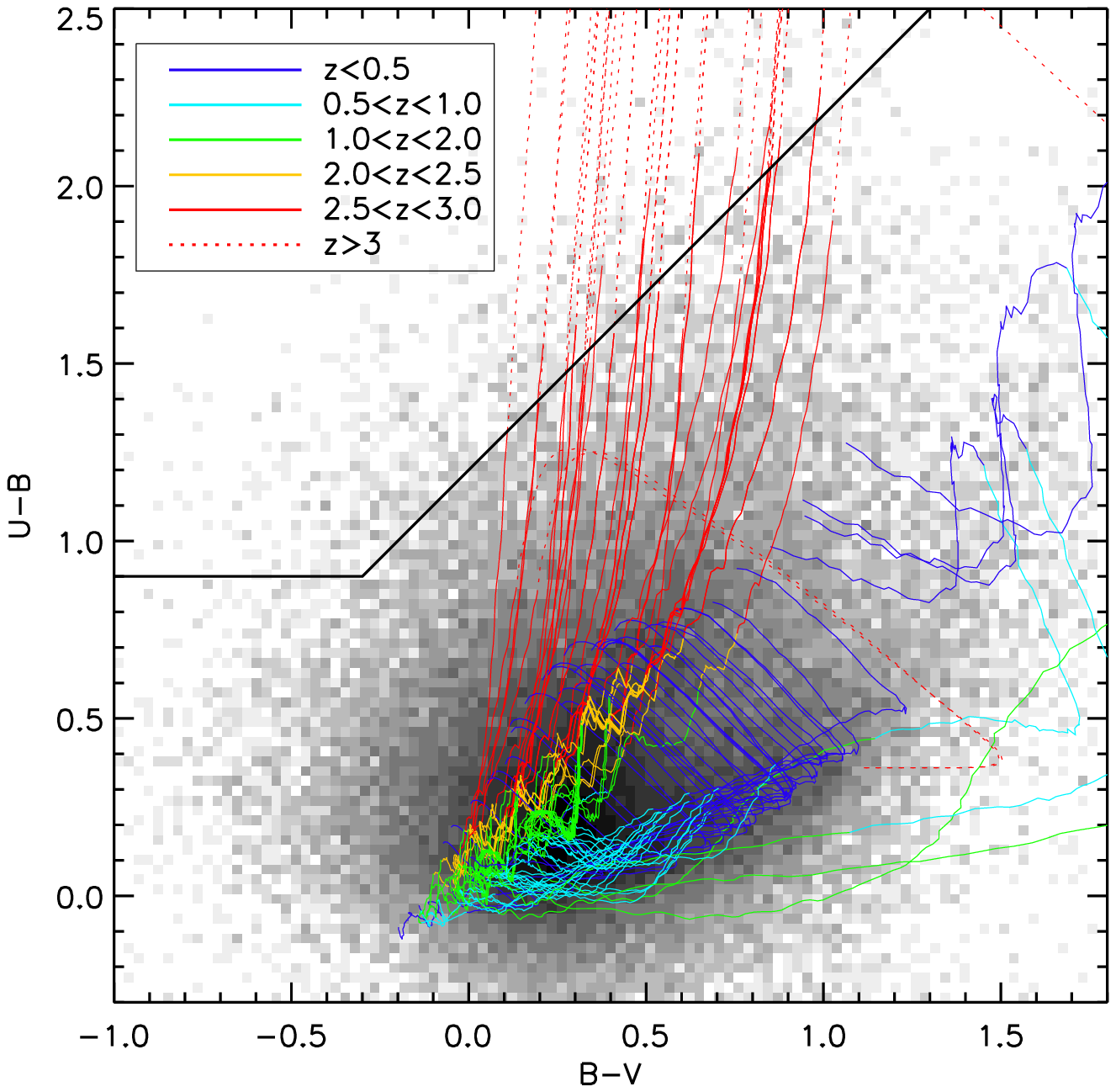}}
  \caption{The same plot as Fig.\,\ref{colours}, but with with SEDs created
           with the GISSEL98 code \citep{Bruzual1993} with different
           metallicity and extinction properties
           \citep[see][]{Bolzonella2000}. We use elliptical, irregular and
           spiral tracks, colour-coded with redshift ranges. The bulk of the
           distribution is reproduced with spiral and irregular tracks at
           redshift $0.5<z<2.5$. The $z\gtrsim3$ region is also marked.
           }
  \label{colours_z}
\end{figure}

At this point it is useful to have a notion about the type of galaxies that
are best represented in our sample and their redshifts. A valuable tool in
this direction is the colour distribution; Fig.\,\ref{colours} shows the
colour plot of the Lockman Hole sources with tracks of templates of different
types of galaxies. We note that the spiral (Sbc-Scd) and irregular tracks lie
closer to the bulk of observed colours. The metallicity and extinction
properties of
a galaxy can have a severe effect in its optical colours. For that reason we
reproduce Fig.\,\ref{colours} with a set of SED templates which have varying
stellar ages, metallicities and extinctions, calculated with the GISSEL98
code \citep{Bruzual1993}. The results are presented in Fig.\,\ref{colours_z},
where the colour tracks are colour-coded with respect to the redshift. The
distribution of sources is well reproduced and we can see that the redshift
range most represented is $z=1.5\pm1$. Moreover, the bulk of the colour
distribution is represented by spiral and irregular tracks, while the
ellipticals account for colours redder than $B-V=1$.

The galaxy types and redshift probed by our survey are compatible with the
``steep faint end slope'' model of \citet{Metcalfe2001} and the slopes are
also in good agreement. Therefore, there is no need to invoke a dwarf galaxy
population to assist the sources which cause the steepening of the faint end
slope, in order to reproduce the B-band data. The U-band counts on the other
hand are underestimated by the ``steep faint end slope'' model, although
the slope itself agrees. In this case a dwarf galaxy population would assist
in incrasing the U-band number counts. It would be however challenging, as
such a population is required to affect the U-band leaving the B-band unchanged.
The Ly-$\alpha$ line falls into the U wavelength range at a redshift of
$z=2$, so a population if highly ionized Ly-$\alpha$ emitters is a good
candidate. However, at $z=2.5$ the blue filter would also be affected, which
leaves a narrow redshift window for this hypothetical population. An
implication of this scenario is a sizeable decrease in the star formation rate
between $z=2$ and $z=2.5$. \citet{Reddy2008} find an increase in the star
formation rate density between $z\sim3$ and $z\sim2$, which is reflected in
the UV luminosity density, and more specifically in the number density of
faint ($M_{\rm AB}(1700\,\AA)>-21$) UV-emitting galaxies.

A steepening of the faint end slope of the (B-band) luminosity function is
already evident since the first computation of its values at redshifts $z>0$
\citep{Lilly1995}. These authors find that the slope increases with redshift
(out to $z=1.3$) and that this is an effect caused by galaxies with blue
optical colours, while the LF of red galaxies shows minimal change in its
fitted parameters. This result is backed up by more recent studies
\citep*{Gabasch2004,Ilbert2005,Arnouts2005,Willmer2006,Prescott2009} and the general trend
is that the not only blue galaxies' LFs evolve more with redshift than those
of redder colours, their faint end slopes are steeper as well. In cases where
the LFs are computed with respect to the galaxy type
\citep{Ilbert2006,Zucca2006}, little (if any) evolution of the faint end slope
is found for each galaxy type, while the slope is different for each type,
the sttepest beeing in irregulars \citep{Zucca2006} or blue-bulge galaxies
\citep{Ilbert2006}. There is however significant change in the normalization
and the value of $M^{\star}$ (the characteristic Schechter luminosity), which
is interpreted as an increase in the fraction of irregular and late-type
galaxies with redshift. The steep faint-end slope we find in the U and B bands
($\alpha\simeq-1.75$) agrees with the values fitted for irregular galaxies
by \citet{Zucca2006} and is even a bit too flat compared to the value assumed
by \citet{Ilbert2005} for the blue-bulge population (their $\alpha=-2.0$).
Given that in this survey the dominant population, especially at faint
magnitudes, is spirals and irregulars at non-local redshifts, we support these
steep faint-end slopes. The sources responsible for the steep slopes are
activly star forming and are good candidates for the ``blue dwarf''
population. \citet{Driver1995} assume that this population consists of
sources with late-type and irregular morphologies; \citet{Ilbert2006} state
that the ``blue bulge'' population could be a population of actively
star-forming galaxies, where the starburst region has bulge-like morphological
characteristics, like the ``blue spheroid'' galaxy sample of \citet{Im2001}.

\begin{figure*}
  \resizebox{\hsize}{!}{\includegraphics{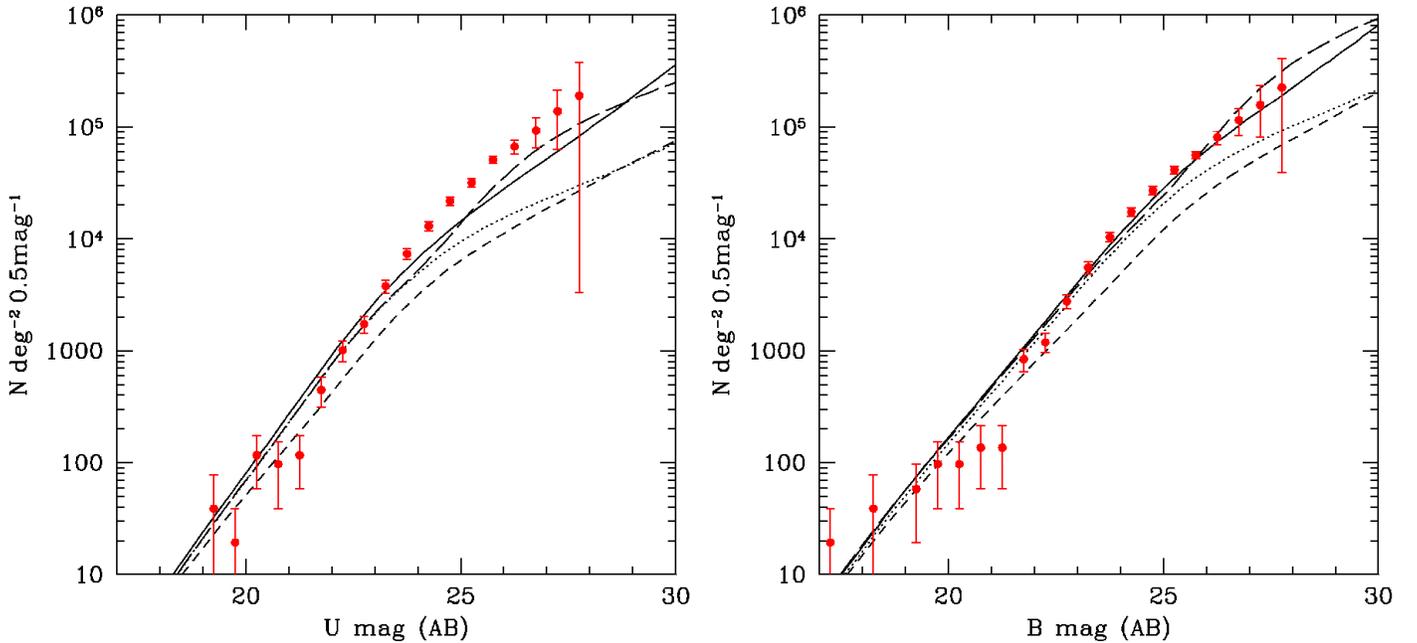}}
  \caption{Measured surface densities in the U and B band (red points),
           plotted with evolution models from \citet{Metcalfe1996} and
           \citet{Metcalfe2001}. The short-dashed lines represent the pure
           luminosity evolution model, the long-dashed lines the same model
           with the inclusion of a population of star-forming dwarf galaxies, 
           and the solid lines the same pure luminosity evolution model with
           a steep faint-end slope of the luminosity function. The latter is
           in very good agreement with the B-band counts, while the U-band
           counts are under-predicted by all models.}
  \label{number_fits_1}
\end{figure*}

\begin{figure*}
  \resizebox{\hsize}{!}{\includegraphics{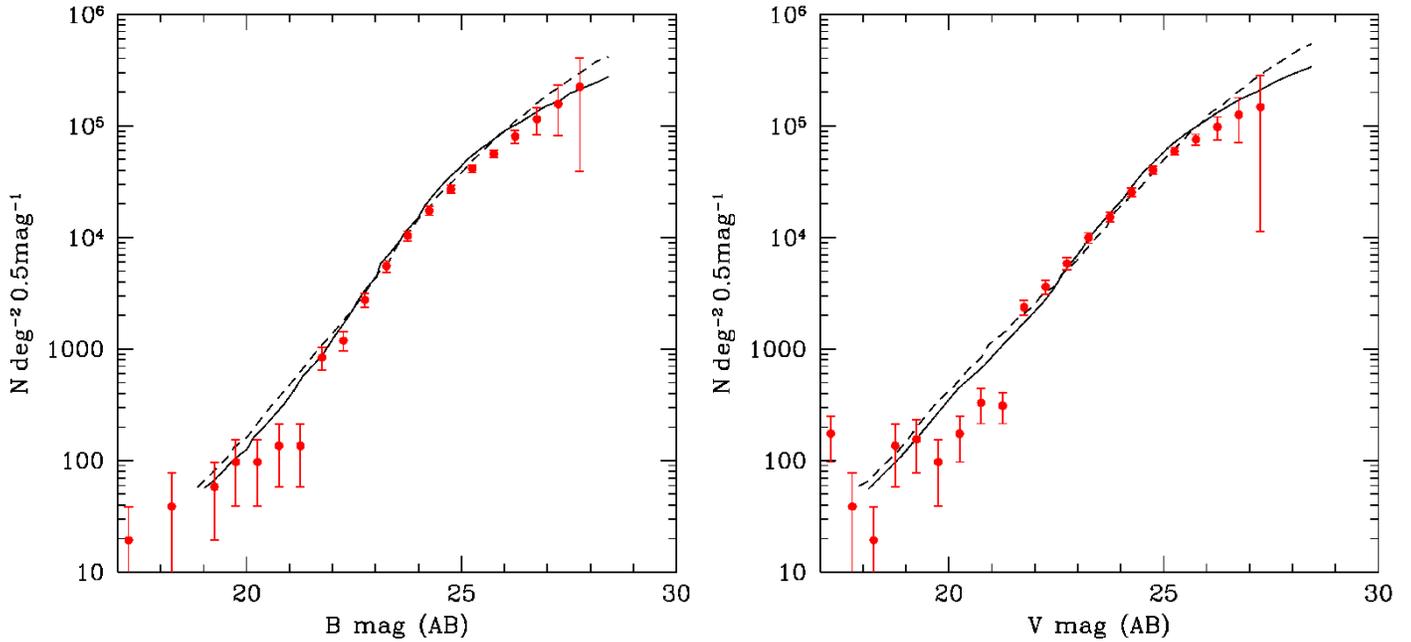}}
  \caption{Measured surface densities in the B and V band (red points),
           plotted with number counts from the simulations of
           \citet{Nagashima2005}. The solid and dashed lines represent the
           strong and weak supernova feedback cases respectively. The
           faint end slope of the weak feedback model is in better agreement
           with the B-band data, while the the strong feedback model is in
           better agreement with the V-band data.}
  \label{number_fits_2}
\end{figure*}

An issue that still needs to be addressed is the flattening of the number
counts slope in the V-band. A mechanism that affects the faint-end slope
is supernova feedback, which is caused by the heating of the interstellar
medium through supernova exposions. \citet{Nagashima2005} have modelled
galaxy formation taking this effect into account. Their predictions for the
number counts using strong or weak feedback (parametrized by the time-scale
in which supernova explosions reheat the cold interstellar gas) differ in the
faint slope with minimal impact on the normalization
\citep[see Figure 18 in][]{Nagashima2005}. Fig.\,\ref{number_fits_2} plots
the B and V-band number counts predictions of \citet{Nagashima2005} with
our data-points; the solid and dashed lines refer to strong and weak
feedback respectively. While both predictions seem to overestimate the
observed number counts at faint fluxes, the faint end slope of the weak
feedback prediction is in good agreement with the data for the B-band, while
the V-band slope is better interpreted with the strong feedback model. A
possible explanation is that the V-band probes the same rest frame wavelength
at higher redshift. If we assume that star formation is the dominant mechanism
producing near-infrared light (where the rest-frame B and V bands are at
redshift $z>1.5$) the V-band probes higher redshifts than the B-band. There
is evidence that the UV luminosity function has a steeper faint end slope
at $z\sim2$ \citep[$\alpha=-1.88\pm0.27$,][]{Reddy2008} than at $z\sim3$
\citep[$\alpha=-1.60\pm0.13$,][]{Steidel1999}. In this case,
starburst feedback would be stronger at higher redshift, in line with our data.
So, enhanced SFR at $z>1.5$ could cause the flattening of the V-band faint
slope. 

\subsection{Colour selection}
\label{udrops}

To be able to test evolutionary models of galaxies in a greater extent one
needs to have information of the redshifts of the various objects found in
a ``blind'' survey. However, even with the largest telescopes available it is
practically impossible to have complete samples beyond $R\sim24$ and use the
full capacity of photometric surveys.
Moreover, the selection of targets for spectroscopy at such faint limits is
hard because their redshift range is so large that it makes it impossible to
get meaningful spectra without pre-selecting the targets according to their
redshift range. A way to overcome this barrier is to use
the photometric redshift
technique, where the SED of each source is compared with known SED templates
to derive an estimate of the redshift. Although the accuracy of this method
is limited so it cannot be used for e.g. spatial clustering studies it can
be very useful in deriving luminosities or selecting objects in different
redshift ranges. Given the extensive spectral coverage of the Lockman Hole
it is possible to calculate photometric redshifts for a large number of
galaxies. Details about the Lockman Hole photo-z survey will be given in a
subsequent paper.

The drawback of the photometric redshift technique is that it requires the
detection of the source in a large number of bands spanning from the near
ultraviolet to the infrared. It is however possible to select sources within
a redshift range using the ``dropout'' technique \citep[e.g.][]{Steidel2003}.
This technique is used to detect the Lyman break in the spectra of galaxies
when it is redshifted between two of the observed bands. In practice the
colour-colour diagram is used to select the sources. Based on the colour
tracks of Fig.\,\ref{colours_z}, we set the selection limits of sources with
$z\gtrsim3$ to:
\[U-B>0.9\]
\[U-B>B-V+1.2\]
\[B-V<1.3\]
Using these limits, we find 2152 sources with redshift $z\gtrsim3$ in the
whole Lockman Hole region (925\,arcmin$^2$), which gives a number density of
8375\,deg$^{-2}$. \citet{Steidel2003} find 6176\,deg$^{-2}$ U-dropout
$z\sim3$ sources using different selection bands ($U_n-G$ vs. $G-\cal R$).
Using the near-infrared coverage of the Lockman Hole we could also select
sources according to their B-z-K colours \citep[see][]{Daddi2004}, having
$1.4<z<2.5$, or according to their i-z colour, having $1.4<z<2.5$
\citep[see][]{Vanzella2005}.
Being able to select sources in distinct redshift ranges provides a valuable
tool to further test models of galaxy evolution.

Selecting sources by their colours can also provide samples of different kinds
of objects. Compton thick AGN are active galaxies with dense environments
($N_{\rm H}>10^{24}\,{\rm cm}^{-2}$) so that they block even hard X-ray
radiation and are not detected even in the deepest X-ray surveys. They would
provide valuable information in evolution studies, as they represent a
distinctive phase of a galaxies lifetime and they are the ``missing link''
in population synthesis models of the X-ray background \citep*{Gilli2007}.
The most promising methods of detecting Compton thick AGN involves comparing
the optical and infrared fluxes of sources
\citep{Donley2007,Daddi2007,Fiore2008,Georgantopoulos2008}. The multi
wavelength coverage of the Lockman Hole in combination with the deep X-ray
observations are ideal for this kind of study.

\subsection{Follow-up}

One important contribution of this study is that it provides a
large number of newly detected extragalactic objects to be further observed
in follow-up campaigns. A number of sources has already been
spectroscopically identified, and they have been selected from the X-ray
campaigns with ROSAT \citep{Schmidt1998,Lehmann2000,Lehmann2001} and
XMM-Newton \citep{Mateos2005}. With multi-object spectrographs we are now
able to conduct spectroscopy to a large number of optical sources. The
LBT is already equipped with a near-infrared multi-slit spectrograph
\citep[LUCIFER;][]{Mandel2007} which will start operation within 2009 and
the optical multi-slit spectrograph \citep[MODS;][]{Pogge2006} is expected
to be operational in 2010. The key scientific goals of these instruments
is to conduct spectroscopy at cosmologically interesting redshifts. 
To be able to select targets for these instruments we need a deep
optical survey and a colour selection scheme similar to what described in the
previous section.


\section{Summary and conclusions}

In this paper we present the deep imaging campaign of the Lockman Hole using
the LBT. The Lockman Hole is an excellent region for deep multi-wavelength
observations given the minimal galactic absorption. Here we report details
of the U, B, and V-band observation and the data reduction strategy.
Our imaging area covers 925\,arcmin$^2$ in a very well sampled region of
the Lockman Hole, with deep X-ray, infrared, and radio coverage.
We have reached depths of 26.7, 26.3, and 26.3 mag(AB) in the U, B, and V
band respectively, in terms of 50\% source detection efficiency, and have
extracted a large number of sources ($\sim 85000$) an all three bands.

The number counts distributions are used to test galaxy evolution models and
and simulations. We find evidence of steepening of the faint-end slope
of the luminosity function in the U and B bands, which can explain the
B number count without the need of a dwarf galaxy population. However the
U counts are under-predicted with this model and an enhancement of the star
formation rate at $z=1.5-2.5$ is needed to explain them. A flatter faint
end slope observed in the V-band case could be the result of supernova
feedback.

This survey is part of an effort to conduct deep observations of the Lockman
Hole in different bands ranging from the infrared to the X-rays. This will
help us select different source classes for further study and in addition to
planned spectroscopic observations create a large database for extragalactic
studies.


\begin{acknowledgements}
The authors thank the LBT Science Demonstration Time (SDT) team for assembling
and executing the SDT program. We also thank the LBC team and the LBTO staff
for their kind assistance.
\end{acknowledgements}


\begin{thebibliography}{}
\scriptsize{
  \bibitem[Alcal\'{a} et al.(2004)]{Alcala2004} Alcal\'{a}, J. M., Pannella, M., Puddu, E., et al., 2004, A\&A, 428, 339
  \bibitem[Arnouts et al.(1997)]{Arnouts1997} Arnouts, S., de Lapparent, V., Mathez, G., Mazure, A., Mellier, Y., Bertin, E., Kruszewski, A., 1997, A\&AS, 124, 163
  \bibitem[Arnouts et al.(1999)]{Arnouts1999} Arnouts, S., D'Odorico, S., Cristiani, S., Zaggia, S., Fontana, A., Giallongo, E., 1999, A\&A, 341, 641
  \bibitem[Arnouts et al.(2001)]{Arnouts2001} Arnouts, S., Vandame, B., Benoist, C., et al., 2001, A\&A, 379, 740
  \bibitem[Arnouts et al.(2005)]{Arnouts2005} Arnouts, S., Schiminovich, D., Ilbert, O., et al., 2005, ApJ, 619L, 43
  \bibitem[Barro et al.(2009)]{Barro2009} Barro, G., Gallego, J., P\'{e}rez-Gonz\'{a}lez, P. G., et al., 2009, A\&A, 494, 63
  \bibitem[Ben\'{i}tez(2000)]{Benitez2000} Ben\'{i}tez, N., 2000, ApJ, 536, 571B
  \bibitem[Berta et al.(2006)]{Berta2006} Berta, S., Rubele, S., Franceschini, A., et al., 2006, A\&A, 451, 881
  \bibitem[Bertin \& Arnouts(1996)]{Bertin1996} Bertin, E., Arnouts, S., 1996, A\&AS, 117, 393
  \bibitem[Bertin \& Dennefeld(1997)]{Bertin1997} Bertin, E., Dennefeld, M., 1997, A\&A, 317, 43
  \bibitem[Biggs \& Ivison(2006)]{Biggs2006} Biggs, A. D., Ivison, R. J., 2006, MNRAS, 371, 963
  \bibitem[Biggs \& Ivison(2008)]{Biggs2008} Biggs, A. D., Ivison, R. J., 2008, MNRAS, 385, 893
  \bibitem[Bolzonella et al.(2000)Bolzonella, Miralles \& Pell\'{o}]{Bolzonella2000} Bolzonella, M., Miralles, J.-M., Pell\'{o}, R., 2000, A\&A, 363, 476
  \bibitem[Brunner et al.(2008)]{Brunner2008} Brunner, H., Cappelluti, N., Hasinger, G., Barcons, X., Fabian, A. C., Mainieri, V., Szokoly, G., 2008, A\&A, 479, 283
  \bibitem[Bruzual \& Charlot(1993)]{Bruzual1993} Bruzual A. G., Charlot, S., 1993, ApJ, 405, 538
  \bibitem[Cabanac et al.(2000)Cabanac, de Lapparent \& Hickson]{Cabanac2000} Cabanac, R. A., de Lapparent, V., Hickson, P., 2000, A\&A, 364, 349
  \bibitem[Capak et al.(2004)]{Capak2004} Capak, P., Cowie, L. L., Hu, E. M., et al., 2004, AJ, 127, 180
  \bibitem[Ciliegi et al.(2003)]{Ciliegi2003} Ciliegi, P., Zamorani, G., Hasinger, G., Lehmann, I., Szokoly, G., Wilson, G., 2003, A\&A, 398, 901
  \bibitem[Coleman et al.(1980)Coleman, Wu \& Weedman]{Coleman1980} Coleman, G. D., Wu, C.-C., Weedman, D. W., 1980, ApJS, 43, 393
  \bibitem[Coppin et al.(2006)]{Coppin2006} Coppin, K., Chapin, E. L., Mortier, A. M. J., et al., 2006, MNRAS, 372, 1621
  \bibitem[Cristiani \& Vio(1990)]{Cristiani1990} Cristiani, S., Vio, R., 1990, A\&A, 227, 385
  \bibitem[Daddi et al.(2004)]{Daddi2004} Daddi, E., Cimatti, A., Renzini, A., Fontana, A., Mignoli, M., Pozzetti, L., Tozzi, P., Zamorani, G., 2004, ApJ, 617, 746
  \bibitem[Daddi et al.(2007)]{Daddi2007} Daddi, E., Alexander, D. M., Dickinson, M., et al., 2007, ApJ, 670, 173
  \bibitem[Donley et al.(2007)]{Donley2007} Donley, J. L., Rieke, G. H., P\'{e}rez-Gonz\'{a}lez, P. G., Rigby, J. R., Alonso-Herrero, A., 2007, ApJ, 660, 167
  \bibitem[Driver et al.(1994)]{Driver1994} Driver, S. P., Phillipps, S., Davies, J. I., Morgan, I., Disney, M. J., 1994, MNRAS, 266, 155
  \bibitem[Driver et al.(1995)]{Driver1995} Driver, S P., Windhorst, R. A., Ostrander, E. J., Keel, W. C., Griffiths, R. E., Ratnatunga, K. U., 1995, ApJ, 449L, 23
  \bibitem[Drory et al.(2001)]{Drory2001} Drory, N., Bender, R., Snigula, J., Feulner, G., Hopp, U., Maraston, C., Hill, G. J., Mendes de Oliveira, C., 2001, ApJ, 562L, 111
  \bibitem[Egami et al.(2008)]{Egami2008} Egami, E., Bock, J., Dole, H., et al., 2008, sptz.prop, 50249E
  \bibitem[Eliche-Moral et al.(2006)]{Eliche-Moral2006} Eliche-Moral, M. C., Balcells, M., Aguerri, J. A. L., Gonz\'{a}lez-Garc\'{i}a, A. C., 2006, ApJ, 639, 644
  \bibitem[Fadda et al.(2004)]{Fadda2004} Fadda, D., Lari, C., Rodighiero, G., Franceschini, A., Elbaz, D., Cesarsky, C., Perez-Fournon, I., 2004, A\&A, 427, 23
  \bibitem[Fiore et al.(2008)]{Fiore2008} Fiore, F., Grazian, A., Santini, P.,  2008, ApJ, 672, 94
  \bibitem[Furusawa et al.(2008)]{Furusawa2008} Furusawa, H., Kosugi, G., Akiyama, M., et al., 2008, ApJS, 176, 1
  \bibitem[Gabasch et al.(2004)]{Gabasch2004} Gabasch, A., Bender, R., Seitz, S., et al., 2004, A\&A, 421, 41
  \bibitem[Gardner et al.(1993)Gardner, Cowie \& Wainscoat]{Gardner1993} Gardner, J. P., Cowie, L. L., Wainscoat, R. J., 1993, ApJ, 415, L9
  \bibitem[Gardner et al.(1996)]{Gardner1996} Gardner, J. P., Sharples, R. M., Carrasco, B. E., Frenk, C. S., 1996, MNRAS, 282L, 1
  \bibitem[Garn et al.(2008)]{Garn2008} Garn, T., Green, D. A., Riley, J. M., Alexander, P., 2008, MNRAS, 387, 1037
  \bibitem[Georgantopoulos et al.(2008)]{Georgantopoulos2008} Georgantopoulos, I., Georgakakis, A., Rowan-Robinson, M., Rovilos, E., 2008, A\&A, 484, 671
  \bibitem[Giallongo et al.(2008)]{Giallongo2008} Giallongo, E., Ragazzoni, R., Grazian, A., et al., 2008, A\&A, 482, 349
  \bibitem[Gilli et al.(2007)Gilli, Comastri \& Hasinger]{Gilli2007} Gilli, R., Comastri, A., Hasinger, G., 2007, A\&A, 463, 79
  \bibitem[Grazian et al.(2009)]{Grazian2009} Grazian, A., Menci, N., Giallongo, E., et al., 2009, A\&A, in press  \texttt{[ArXiv:astro-ph/0906.4035]}
  \bibitem[Greve et al.(2004)]{Greve2004} Greve, T. R., Ivison, R. J., Bertoldi, F., Stevens, J. A., Dunlop, J. S., Lutz, D., Carilli, C. L., et al., 2004, MNRAS, 354, 779
  \bibitem[Guhathakurta et al.(1990)]{Guhathakurta1990} Guhathakurta, P., Tyson, J. A., Majewski, S. R., 1990, in Evolution of the universe of galaxies, Astronomical Society of the Pacific, 304
  \bibitem[Hasinger et al.(1998)]{Hasinger1998} Hasinger, G., Burg, R., Giacconi, R., Schmidt, M., Trumper, J., Zamorani, G., 1998, A\&A, 329, 482
  \bibitem[Hasinger et al.(2001)]{Hasinger2001} Hasinger, G., Altieri, B., Arnaud, M., et al., 2001, A\&A, 365, L45
  \bibitem[Heydon-Dumbleton et al.(1989)Heydon-Dumbleton, Collins \& MacGillivray]{Heydon-Dumbleton1989} Heydon-Dumbleton, N. H., Collins, C. A., MacGillivray, H. T., 1989, MNRAS, 238, 379
  \bibitem[Hogg et al.(1997)]{Hogg1997} Hogg, D. W., Pahre, M. A., McCarthy, J. K., Cohen, J. G., Blandford, R., Smail, I., Soifer, B. T., 1997, MNRAS, 288, 404
  \bibitem[Hopkins(2004)]{Hopkins2004} Hopkins, A. M., 2004, ApJ, 615, 209
  \bibitem[Huang et al.(2001)]{Huang2001} Huang, J.-S., Thompson, D., K\"{u}mmel, M. W., et al., 2001, A\&A, 368, 787
  \bibitem[Huang et al.(2004)]{Huang2004} Huang, J.-S., Barmby, P., Fazio, G. G., et al., 2004, ApJS, 154, 44
  \bibitem[Ilbert et al.(2005)]{Ilbert2005} Ilbert, O., Tresse, L., Zucca, E., et al., 2005, A\&A, 439, 863
  \bibitem[Ilbert et al.(2006)]{Ilbert2006} Ilbert, O., Lauger, S., Tresse, L., et al., 2006, A\&A, 453, 809
  \bibitem[Ilbert et al.(2009)]{Ilbert2009} Ilbert, O., Capak, P., Salvato, M., 2009, ApJ, 690, 1236
  \bibitem[Irwin et al.(1994) Irwin, Maddox \& McMahon]{Irwin1994}  Irwin, M., Maddox, S., McMahon, R. G., 1994, Spectrum, 2, 14
  \bibitem[Im et al.(2001)]{Im2001} Im, M., Faber, S. M., Gebhardt, K., Koo, D. C., Phillips, A. C., Schiavon, R. P., Simard, L., Willmer, C. N. A., 2001, AJ, 122, 750
  \bibitem[Ivison et al.(2002)]{Ivison2002} Ivison, R. J., Greve, T. R., Smail, I., et al., 2002, MNRAS, 337, 1
  \bibitem[Jarrett et al.(1994)Jarrett, Dickman \& Herbst]{Jarrett1994} Jarrett, T. H., Dickman, R. L., Herbst, W., 1994, ApJ, 424, 852
  \bibitem[Jones et al.(1991)]{Jones1991} Jones, L. R., Fong, R., Shanks, T., Ellis, R. S., Peterson, B. A., 1991, MNRAS, 249, 481

  \bibitem[Kaiser et al.(1999)]{Kaiser1999} Kaiser, N., Wilson, G., Luppino, G., Dahle, H., 1999, PASP, submitted \texttt{[ArXiv:astro-ph/9907.229]}
  \bibitem[Kashikawa et al.(2004)]{Kashikawa2004} Kashikawa, N., Shimasaku, K., Yasuda, N., et al., 2004, PASJ, 56, 1011
  \bibitem[Kawara et al.(2004)]{Kawara2004} Kawara, K., Matsuhara, H., Okuda, H., et al., 2004, A\&A, 413, 843
  \bibitem[Koo(1986)]{Koo1986} Koo, D., 1986, ApJ, 311, 651
  \bibitem[K\"{u}mmel et al.(2001)]{Kummel2001} K\"{u}mmel, M. W., Wagner, S. J., 2001, A\&A, 370, 384
  \bibitem[Landolt(1992)]{Landolt1992} Landolt, A. U., 1992, AJ, 104, 372
  \bibitem[Lawrence et al.(2007)]{Lawrence2007} Lawrence, A., Warren, S. J., Almaini, O., et al., 2007, MNRAS, 379, 1599
  \bibitem[Laurent et al.(2005)]{Laurent2005} Laurent, G. T., Aguirre, J. E., Glenn, J., et al., 2005, ApJ, 623, 742
  \bibitem[Lehmann et al.(2000)]{Lehmann2000} Lehmann, I., Hasinger, G., Schmidt, M., et al., 2000, A\&A, 354, 35
  \bibitem[Lehmann et al.(2001)]{Lehmann2001} Lehmann, I., Hasinger, G., Schmidt, M., et al., 2001, A\&A, 371, 833
  \bibitem[Lilly et al.(1991)Lilly, Cowie \& Gardner]{Lilly1991} Lilly, S. J., Cowie, L. L., Gardner, J. P., 1991, ApJ, 369, 79
  \bibitem[Lilly et al.(1995)]{Lilly1995} Lilly, S. J., Tresse, L., Hammer, F., Crampton, D., Le F\`{e}vre, O., 1995, ApJ, 155, 108
  \bibitem[Liske et al.(2003)]{Liske2003} Liske, J., Lemon, D. J., Driver, S. P., Cross, N. J. G., Couch, W. J., 2003, MNRAS, 344, 307
  \bibitem[Lockman et al.(1986)Lockman, Jahoda \& McCammon]{Lockman1986} Lockman, F. J., Jahoda, K., McCammon, D., 1986, ApJ, 302, 432
  \bibitem[Lonsdale et al.(2003)]{Lonsdale2003} Lonsdale, C. J., Smith, H. E., Rowan-Robinson, M., et al., 2003, PASP, 115, 897
  \bibitem[Maddox et al.(1990)]{Maddox1990} Maddox, S. J., Sutherland, W. J., Efstathiou, G., Loveday, J., Peterson, B. A., 1990, MNRAS, 247, 1
  \bibitem[Mandel et al.(2007)]{Mandel2007} Mandel, H., Seifert, W., Lenzen, R., et al., 2007, AN, 328, 626
  \bibitem[Martin et al.(2005)]{Martin2005} Martin, D. C., Fanson, J., Schiminovich, D., et al., 2005, ApJ, 619, L1
  \bibitem[Mateos et al.(2005)]{Mateos2005} Mateos, S., Barcons, X., Carrera, F. J., Ceballos, M. T., Hasinger, G., Lehmann, I., Fabian, A. C., Streblyanska, A., 2005, A\&A, 444, 79
  \bibitem[McCracken et al.(2001)]{McCracken2001} McCracken, H. J., Le F\`{e}vre, O., Brodwin, M., Foucaud, S., Lilly, S. J., Crampton, D., Mellier, Y., 2001, A\&A, 376, 756
  \bibitem[McCracken et al.(2003)McCracken, Radovich \& M., Bertin]{McCracken2003} McCracken, H. J., Radovich, M., Bertin, E., 2003, A\&A, 410, 17
  \bibitem[Metcalfe et al.(1991)]{Metcalfe1991} Metcalfe, N., Shanks, T., Fong, R., Jones, L. R., 1991, MNRAS, 249, 498
  \bibitem[Metcalfe et al.(1995)]{Metcalfe1995} Metcalfe, N., Shanks, T., Fong, R., Roche, N., 1995, MNRAS, 273, 257
  \bibitem[Metcalfe et al.(1996)]{Metcalfe1996} Metcalfe, N., Shanks, T., Campos, A., Fong, R., Gardner, J. P., 1996, Nature, 383, 236
  \bibitem[Metcalfe et al.(2001)]{Metcalfe2001} Metcalfe, N., Shanks, T., Campos, A., McCracken, H. J., Fong, R., 2001, MNRAS, 323, 795
  \bibitem[Monet(1998)]{Monet1998} Monet, D. G., 1998, AAS, 19312003
  \bibitem[Nagashima et al.(2005)]{Nagashima2005} Nagashima, M., Yahagi, H., Enoki, M., Yoshii, Y., Gouda, N., 2005, ApJ, 634, 26
  \bibitem[Pogge et al.(2006)]{Pogge2006} Pogge, R. W., Atwood, B., Belville, S. R, et al., 2006, SPIE, 6269, 16
  \bibitem[Prandoni et al.(1999)Prandoni, Wichmann \& da Costa]{Prandoni1999} Prandoni, I., Wichmann, R., da Costa, L., et al., 1999, A\&A, 345, 448
  \bibitem[Prescott et al.(2009)Prescott, Baldry \& James]{Prescott2009} Prescott, M., Baldry, I. K., James, P. A., 2009, MNRAS, 397, 90
  \bibitem[Reddy et al.(2008)]{Reddy2008} Reddy, N. A., Steidel, C. C., Pettini, M., Adelberger, K. L., Shapley, A. E., Erb, D. K., Dickinson, M., 2008, ApJS, 175, 48
  \bibitem[Rodighiero et al.(2004)]{Rodighiero2004} Rodighiero, G., Lari, C., Fadda, D., Franceschini, A., Elbaz, D., Cesarsky, C., 2004, A\&A, 427, 773
  \bibitem[Schmidt et al.(1998)]{Schmidt1998} Schmidt, M., Hasinger, G., Gunn, J., et al., 1998, A\&A, 329, 495
  \bibitem[Scott et al.(2006)]{Scott2006} Scott, K., et al., 2006, AAS, 209, 8303
  \bibitem[Smail et al.(1995)]{Smail1995} Smail, I., Hogg, D. W., Yan, L., Cohen, J. G., 1995, ApJ, 449L, 105
  \bibitem[Sommerville et al.(2004)]{Sommerville2004} Sommerville, R. S., Lee, K., Ferguson H. C., Gardner, J. P., Moustakas, L. A., Giavalisco, M., 2004, ApJ, 600L, 171
  \bibitem[Songaila et al.(1990)Songaila, Cowie \& Lilly]{Songaila1990} Songaila, A., Cowie, L. L., Lilly, S. J., 1990, ApJ, 348, 371
  \bibitem[Steidel et al.(1999)]{Steidel1999} Steidel, C. C.; Adelberger, K. L., Giavalisco, M., Dickinson, M., Pettini, M., 1999, ApJ, 519, 1
  \bibitem[Steidel et al.(2003)]{Steidel2003} Steidel, C. C., Adelberger, K. L., Shapley, A. E., Pettini, M., Dickinson, M., Giavalisco, M., 2003, ApJ, 592, 728
  \bibitem[Radovich et al.(2004)]{Radovich2004} Radovich, M., Arnaboldi, M., Ripepi, V., et al., 2004, A\&A, 417, 51
  \bibitem[Somerville et al.(2008)]{Somerville2008} Somerville, R. S., Hopkins, P. F., Cox, T. J., Robertson, B. E., Hernquist, L., 2008, MNRAS, 391, 481
  \bibitem[Szalay et al.(1999)Szalay, Connolly \& Szokoly]{Szalay1999} Szalay, A. S., Connolly, A. J., Szokoly, G. P., 1999, AJ, 117, 68
  \bibitem[Trenti \& Stiavelli(2008)]{Trenti2008} Trenti, M., Stiavelli, M., 2008, ApJ, 676, 767
  \bibitem[Tyson(1988)]{Tyson1988} Tyson, J. A., 1988, AJ, 96, 1
  \bibitem[Ueda et al.(2003)]{Ueda2003} Ueda, Y., Akiyama, M., Ohta, K., Miyaji, T., 2003, ApJ, 598, 886
  \bibitem[van den Bergh(2001)]{vandenBergh2001} van den Bergh, S., 2001, AJ, 122..621
  \bibitem[Vanzella et al.(2005)]{Vanzella2005} Vanzella, E., Cristiani, S., Dickinson, M., et al., 2005, A\&A, 434, 53
  \bibitem[Vollmer et al.(2008)Vollmer, Beckert \& Davies]{Vollmer2008} Vollmer, B., Beckert, T., Davies, R. I., 2008, A\&A, 491, 441
  \bibitem[Volonteri et al.(2000)]{Volonteri2000} Volonteri, M., Saracco, P., Chincarini, G., Bolzonella, M., 2000, A\&A, 362, 487
  \bibitem[Wadadekar et al.(2006)Wadadekar, Casertano \& de Mello]{Wadadekar2006} Wadadekar, Y., Casertano, S., de Mello, D., 2006, ApJ, 123, 1023
  \bibitem[Williams et al.(1996)]{Williams1996} Williams, R. E., Blacker, B., Dickinson, M., et al., 1996, AJ, 112, 1335
  \bibitem[Willmer et al.(2006)]{Willmer2006} Willmer, C. N. A., Faber, S. M., Koo, D. C., et al., 2006, ApJ, 647, 583
  \bibitem[Wilson et al.(2001)]{Wilson2001} Wilson, G., Kaiser, N., Luppino, G. A.; Cowie, L. L., 2001, ApJ, 555, 572
  \bibitem[Wilson(2003)]{Wilson2003} Wilson, G., 2003, ApJ, 585, 191
  \bibitem[Yasuda et al.(2001)]{Yasuda2001} Yasuda, N., Fukugita, M., Narayanan, V. K., et al., 2001, AJ, 122, 1104
  \bibitem[Zucca et al.(2006)]{Zucca2006} Zucca, E., Ilbert, O., Bardelli, S., et al., 2006, A\&A, 455, 879
}
\end{thebibliography}
\end{document}